\pdfoutput=1
\documentclass[11pt,a4paper]{article}

\usepackage[T1]{fontenc}
\usepackage[utf8]{inputenc}
\usepackage{lmodern}
\usepackage{microtype}
\usepackage[margin=22mm]{geometry}
\usepackage{amsmath,amssymb,amsthm,mathtools}
\usepackage{booktabs,tabularx,array}
\usepackage{graphicx}
\usepackage{xcolor}
\usepackage[authoryear,round]{natbib}
\usepackage[colorlinks=true,linkcolor=navy,citecolor=navy,urlcolor=navy]{hyperref}

\definecolor{navy}{HTML}{1B365D}
\definecolor{slate}{HTML}{2D3748}

\theoremstyle{plain}
\newtheorem{theorem}{Theorem}[section]
\newtheorem{lemma}{Lemma}[section]

\newtheorem{proposition}{Proposition}[section]

\title{%
  \textbf{\Large Biophysical Scaling Laws of Radiation Cell Death:}\\[3pt]
  \textbf{\large Chromosome-Arm Targets, Genome-Size Amplification,}\\[1pt]
  \textbf{\large Apparent Biphasic Recovery, and Multiscale Hierarchy}%
}

\author{%
  Yuichiro Manabe$^{1,*}$, \quad Yukitoshi Hirota$^{1}$, \quad Yoko Akiyama$^{1}$, \quad Fuminobu Sato$^{1}$\\[4pt]
  {\small $^{1}$Division of Sustainable Energy and Environmental Engineering, Graduate School of Engineering,}\\
  {\small The University of Osaka, 2-1 Yamadaoka, Suita, Osaka 565-0871, Japan}\\
  {\small $^{*}$Corresponding author: \texttt{manabe\_y@see.eng.osaka-u.ac.jp}}
}
\date{November 27, 2026\vspace{-8pt}}

\begin{document}
\maketitle

\begin{abstract}
\citet{shibata2026dynamic} (\textit{J.\ Theor.\ Biol.}\ 637, 112620) formulated a dynamic ordinary differential equation (ODE) system incorporating first-order sub-lethal damage (SLD) recovery into the classical multi-target single-hit cell-killing model, reproducing the split-dose V79 cell survival experiments of \citet{elkind1960radiation} with a single parameter set (target number $n=5$, damage induction rate $\alpha=0.716\text{ Gy}^{-1}$, and recovery rate $\beta=0.00596\text{ min}^{-1}$). Here, by reformulating this ODE system analytically without introducing any additional fitting parameters, we present a theoretical framework that may coherently interpret four long-standing questions in radiobiology (calibrated primarily against V79 data and published model parameters).

First, rigorously correcting for redundant hits onto already-damaged targets---which occur an average of $nH_5 = 11.42$ times before lethality (hit efficiency $1/H_5 \approx 43.8\%$)---yields an effective DNA target length of $L_1 \approx 90\text{--}140\text{ Mbp}$ per target ($9.0\text{--}17.9\%$ of the genome overall), broadly consistent with a single \textbf{chromosome arm ($\sim 109\text{ Mbp}$)} in V79 cells. This physical correspondence yields a testable prediction: the dependence of the terminal survival slope $1/D_0$ on genome size $G$ is amplified from the classical high-dose-rate Sparrow scaling ($\propto G^1$) to a \textbf{$G^n$ scaling law} (up to a $32$-fold difference for $n=5$ when $G$ doubles) at low dose rates below the characteristic dose rate $\gamma_* = \beta/\alpha \approx 0.50\text{ Gy/h}$. Second, because the quasi-threshold dose $D_q(\tau)$ obeys the exact logarithmic identity $D_q(\tau) = \alpha^{-1}\ln[U(\tau)/S(D_1)]$ in terms of the effective intact target count $U(\tau)$, even a \textbf{purely single recovery rate $\beta$ (half-time $116.3\text{ min}$) can mathematically generate a $2.40$-fold apparent initial acceleration and apparent biphasic recovery ($27.1\text{ min}$, $35.4\%$ $+$ $106.5\text{ min}$, $64.6\%$)} via logarithmic concavity; we also derive a linearizing diagnostic to discriminate true two-pathway repair from logarithmic curvature. Third, showing that the Shibata model extends the Whack-A-Mole (WAM) model of \citet{manabe2015whack} (a single time-ODE balancing damage induction and repair) into a multi-stage ODE hierarchy demonstrates how processes from $\mathrm{SSB}\to\mathrm{DSB}$ formation ($n=2$, genome-length-invariant crossover dose $17.5\text{--}35\text{ kGy}$) through cell death ($\sim 100\text{ Mbp}$, $2.80\text{ h}$) to mouse stem-cell mutagenesis ($41.7\text{ days}$) can be described within a single mathematical skeleton separated cleanly by spatial ($\sim 10^7$-fold) and temporal ($\sim 358$-fold) scales. Finally, we prove within this framework that the transient survival dip in Elkind fractionation originates from cell-cycle redistribution $\alpha(\tau)$ rather than non-monotonic $\beta(t)$, and provide a compact matrix-entry formulation for post-recovery proliferation with colony multiplicity. All rigorous theorems, propositions, proofs, and numerical verification tables are collected in Appendices~A--E so that the main text focuses on biophysical intuition and experimental tests.
\end{abstract}


% =========================================================================
\section{Introduction: Four Open Questions in Radiobiology}
\label{sec:intro}
% =========================================================================

Quantitative understanding of low-LET radiation effects across molecular, cellular, and tissue scales is essential both for high-precision radiotherapy and for radiation protection under low-dose-rate exposures. Despite more than half a century of experimental and theoretical progress, four fundamental questions in radiobiology have largely been treated as disconnected phenomena:

\begin{enumerate}
  \item \textbf{What is the physical identity of the ``target'' of radiation cell death?} Since the multi-target framework of \citet{elkind1959xray,elkind1960radiation}, effective target numbers $n \approx 5$ and mean inactivation doses $M$ have been extracted from survival curve shoulders, yet ambiguity persisted as to whether a target corresponds to individual genes (kilobase scale) or higher-order chromosomal architecture.
  \item \textbf{How does genome size couple to the dose-rate effect?} At high dose rates (HDR), cellular radiosensitivity $1/D_0$ empirically scales with nuclear DNA content or chromosome volume (Sparrow's law \citep{sparrow1967chromosomes}). However, a unified prediction of how differences in genome size propagate into low-dose-rate (LDR) radiosensitivity---distinct from ploidy effects (haploid vs.\ diploid)---has been lacking.
  \item \textbf{Is ``biphasic recovery'' in split-dose curves always evidence for two repair pathways (NHEJ + HR)?} Split-dose recovery curves of the quasi-threshold dose $D_q(\tau)$ typically rise steeply over $1\text{--}2\text{ h}$ before saturating slowly. Two-exponential fits to such data are widely interpreted as direct evidence for fast non-homologous end joining (NHEJ) plus slow homologous recombination (HR), raising the question of whether nonlinear distortion is already inherent in macroscopic $D_q(\tau)$ itself.
  \item \textbf{Why is DSB induction linear while cell killing and mutagenesis are multi-hit and dose-rate-dependent?} At standard doses (several Gy), DNA double-strand break (DSB) yields scale nearly linearly with dose and remain largely dose-rate-independent, whereas cell killing and tissue mutagenesis display strong multi-hit curvature and dose-rate sparing. Likewise, debate has persisted on whether the transient survival reversal (``dip'') around $\tau = 4\text{--}6\text{ h}$ in \citet{elkind1960radiation} Fig.~16 reflects temporary repair suppression or cell-cycle redistribution.
\end{enumerate}

Recently, \citet{shibata2026dynamic} (hereafter the ``Shibata model'') drew upon the dynamic-equilibrium Whack-A-Mole (WAM) concept of \citet{manabe2015whack} to build a coupled ODE system tracking damaged target states $i = 0,1,\dots,n$ with first-order recovery transitions ($i \to i-1$), accurately reproducing the \citet{elkind1960radiation} V79 split-dose dataset ($\tau = 0, 2.5, 23\text{ h}$) using a single parameter set ($n=5$, $\alpha = 0.716\text{ Gy}^{-1}$, $\beta = 0.00596\text{ min}^{-1}$). Because that study relied primarily on numerical integration under specific experimental protocols, the general scaling laws embedded in the system and their connection to the four questions above remained unexplored.

The central finding of the present work lies in reformulating the Shibata model analytically. Whereas the 2015 WAM model consisted of a single time-ODE balancing physical damage induction against biological repair/clearance to determine mutation frequency, the Shibata model preserves that core competition while extending it into a multi-stage coupled ODE system terminating in an irreversible cell-death state ($i=n$). This analytical reformulation reveals that the Shibata model is a natural multi-stage generalization of WAM. Using closed-form solutions and exact spectral bounds without adding free fitting parameters, we find clear quantitative links between the already-calibrated model parameters and established biophysical data for all four open questions, yielding falsifiable experimental predictions. Because our primary quantitative calibration rests on V79 cell survival data and previously published WAM parameters, systematically testing the generality of these predictions across diverse cell lines and organisms remains an important direction for future work.

\paragraph{Organization of This Paper}
To keep the main text (Sections~\ref{sec:framework}--\ref{sec:conclusion}) concise and accessible to radiobiologists, matrix algebra and theorem proofs are separated from physical intuition and biological tests. Section~\ref{sec:framework} explains the conceptual relation between the 2015 WAM model and the multi-stage Shibata extension, showing that the split-dose numerical results of \citet{shibata2026dynamic} reduce to exact closed-form expressions and identities. Sections~\ref{sec:finding_target_genome}--\ref{sec:finding_multiscale} develop four biophysical results implied by these solutions:
\begin{itemize}
  \item the single-target physical size directly tied to $\alpha$ matches the V79 chromosome-arm scale ($\sim 100\text{ Mbp}$) once redundant hits are corrected (Section~\ref{subsec:target_arm}),
  \item the terminal survival slope $1/D_0$ (smallest matrix eigenvalue $\mu_1$) scales as $G^1$ at HDR but amplifies to $G^n$ at LDR (Section~\ref{subsec:genome_scaling}),
  \item even a single biological repair constant $\beta$ naturally generates an apparent fast-plus-slow biphasic $D_q(\tau)$ curve via $D_q \propto \ln U$, alongside two diagnostic tests to separate real two-pathway kinetics from logarithmic distortion (Section~\ref{sec:finding_biphasic}), and
  \item because $\alpha$ scales with target length, $\mathrm{SSB}\to\mathrm{DSB}$ coincidence ($\sim 10\text{ bp}$), cellular lethality ($\sim 100\text{ Mbp}$), and stem-cell mutagenesis ($\sim 42\text{ days}$) occupy a single mathematical skeleton separated by space and time scales (Section~\ref{sec:finding_multiscale}).
\end{itemize}
Section~\ref{sec:implications} proves that cell-cycle sensitivity modulation $\alpha(\tau)$---not $\beta(t)$---causes the split-dose survival dip, formulates G2-checkpoint-gated proliferation with colony multiplicity via a single matrix-entry modification, and summarizes experimental tests and model limitations. All rigorous theorems, propositions, proofs, and numerical verification tables appear in Appendices~A--E.

% =========================================================================
\section{Model Structure: Multi-Stage Extension of WAM and Exact Split-Dose Solutions}
\label{sec:framework}
% =========================================================================

\subsection{From the 2015 WAM Model to the Multi-Stage Shibata Model}

The fate of irradiated cells is governed by the competition between two rate processes: physical damage induction on intact targets (rate $\alpha\ [\mathrm{Gy}^{-1}]$ per target per unit dose, i.e., time rate $\alpha\gamma$ at dose rate $\gamma$) and biological recovery of damaged targets (rate $\beta\ [\mathrm{min}^{-1}]$ per damaged target per unit time; denoted $\kappa$ and $\rho$ in \citet{shibata2026dynamic}).

\begin{figure}[tbp]
\centering
\includegraphics[width=0.98\textwidth]{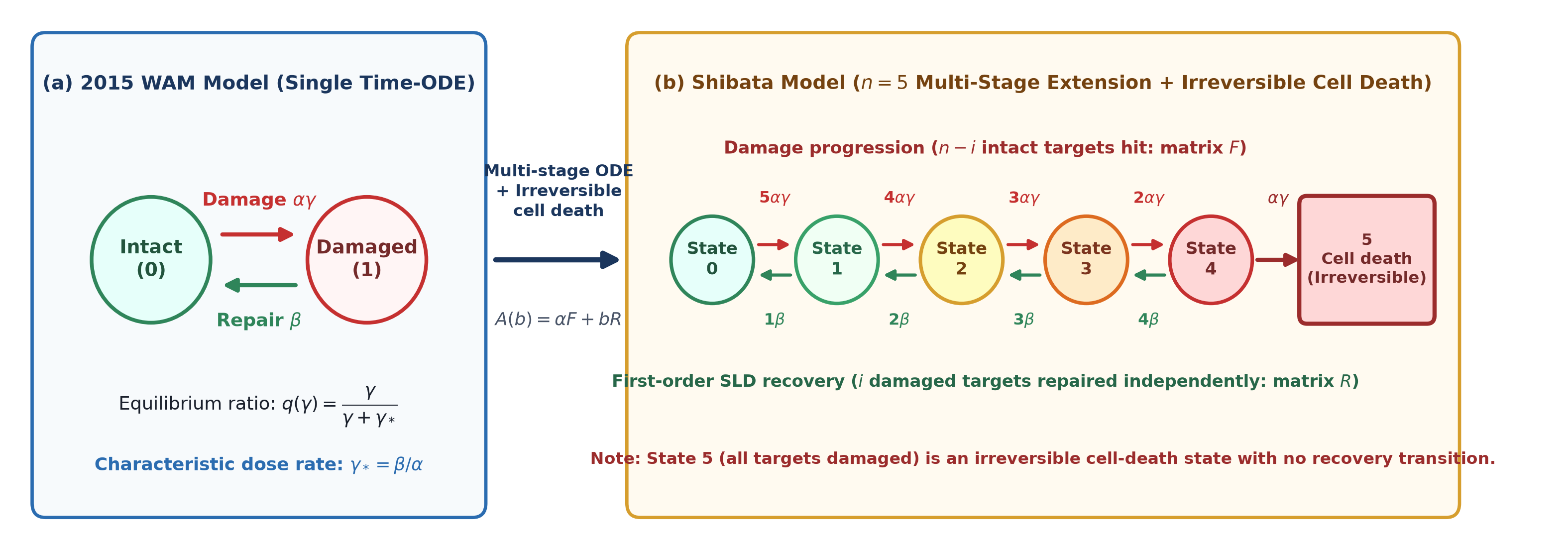}
\caption{Physical correspondence between the 2015 WAM model and the Shibata model. (a) \citet{manabe2015whack} WAM model: a single time-ODE balancing damage induction $\alpha\gamma$ and repair $\beta$ between intact (0) and damaged (1) states. (b) \citet{shibata2026dynamic} multi-target cell-death model ($n=5$): a multi-stage ODE extension terminating in irreversible cell death (state 5), compactly decomposed as $A(b)=\alpha F + bR$ with $b=\beta/\gamma$.}
\label{fig:wam_schematic}
\end{figure}

The WAM model introduced by \citet{manabe2015whack} for radiation mutagenesis describes the balance between physical damage induction and biological repair/elimination using a single time-ODE (Fig.~\ref{fig:wam_schematic}(a)). Carrying over this core mechanism, the Shibata model generalizes it to a multi-stage coupled ODE tracking $i=0,1,\dots,n-1$ damaged targets among $n$ independent targets, assuming that \textbf{once all $n$ targets are damaged simultaneously (state $n$), the cell enters an irreversible death state from which no recovery transition exists} (Fig.~\ref{fig:wam_schematic}(b)).

Changing the independent variable from time $t$ to accumulated dose $D = \gamma t$ decomposes the ODE system cleanly into the sum of two fixed integer $n\times n$ matrices determined solely by $n$---a lower bidiagonal damage-progression matrix $F$ (Forward) and an upper bidiagonal repair matrix $R$ (Recovery):
\begin{equation}
\frac{d\mathbf{p}}{dD} = -A(b)\,\mathbf{p}, \qquad A(b) = \alpha F + b R, \qquad b = \frac{\beta}{\gamma} \ge 0
\label{eq:main_matrix}
\end{equation}
(component definitions, unified nomenclature, and properties are detailed in Appendix~A). Balancing the forward coefficient $\alpha$ against the per-dose recovery coefficient $b=\beta/\gamma$ defines the natural \textbf{characteristic dose rate}:
\begin{equation}
\gamma_* \equiv \frac{\beta}{\alpha},
\label{eq:gamma_star_def}
\end{equation}
which separates the HDR regime ($\gamma \gg \gamma_*$, negligible repair during irradiation) from the LDR regime ($\gamma \ll \gamma_*$, repair outpacing hit accumulation). For the V79 parameter set calibrated in \citet{shibata2026dynamic} ($n=5$, $\alpha=0.716\text{ Gy}^{-1}$, $\beta=0.00596\text{ min}^{-1}$), we find $\gamma_* = 0.008324\text{ Gy/min} \approx 0.50\text{ Gy/h}$, squarely matching the clinical low-dose-rate brachytherapy window ($0.4\text{--}2.0\text{ Gy/h}$ \citep{icru1985report38}).

Note that whereas the 2015 WAM model explicitly included a baseline spontaneous mutation rate ($a_0$) under unirradiated conditions, the present cell-death formulation assumes normalization by the unirradiated plating efficiency in short-term clonogenic assays and omits \textbf{spontaneous / background cell death} due to endogenous damage or baseline apoptosis (see also Section~\ref{sec:implications}).

\subsection{Exact Closed-Form Split-Dose Solution and Universal 95\% Recovery Time}

In an acute split-dose protocol (in the Elkind experiments, $\gamma = 7.22\text{ Gy/min} \approx 867\,\gamma_*$) with first dose $D_1$, recovery interval $\tau$, and second dose $d$, each sub-lethal lesion created by $D_1$ survives the recovery period independently with probability $r = e^{-\beta\tau}$ (Lemma~C.1). Without numerical ODE integration, the surviving fraction reduces to an exact elementary closed form (Theorem~C.1):
\begin{equation}
S(D_1,\tau,d) = 1 - \bigl[q_1 r + q_2(1-q_1 r)\bigr]^n - q_1^n\Bigl(1 - \bigl[r + q_2(1-r)\bigr]^n\Bigr),
\label{eq:main_split}
\end{equation}
where $q_1 = 1 - e^{-\alpha D_1}$ and $q_2 = 1 - e^{-\alpha d}$. The third term removes the fraction $q_1^n$ of cells killed immediately during $D_1$ so that irreversible death cannot participate in recovery. Equation~\eqref{eq:main_split} reproduces all 22 official V79 experimental points of \citet{elkind1960radiation} and \citet{shibata2026dynamic} to machine precision ($R_{\mathrm{adj}}^2 = 0.9858$, Fig.~\ref{fig:split_dq}(a); full verification in Appendix~A Table~A.2). The upward deviation of the $23\text{ h}$ experimental points reflects post-recovery cell proliferation (addressed in Section~\ref{sec:implications}).

Moreover, the quasi-threshold dose $D_q(\tau)$---the dose-axis intercept of the high-dose exponential tail of $S(D_1,\tau,d)/S(D_1)$ measuring shoulder restoration---and the effective intact target pool among survivors, $U(T_1+\tau) = \sum_{i=0}^{n-1}(n-i)p^{(i)}(T_1+\tau)$, satisfy the \textbf{exact identity} (Theorem~C.2):
\begin{equation}
D_q(\tau) = \frac{1}{\alpha}\ln\!\left(\frac{U(T_1+\tau)}{S(D_1)}\right), \qquad
U(T_1+\tau) = n S(D_1) - \bigl[n S(D_1) - n e^{-\alpha D_1}\bigr]e^{-\beta\tau}.
\label{eq:main_Dq}
\end{equation}
Although both the initial value $U(T_1) = n e^{-\alpha D_1}$ and asymptote $n S(D_1)$ depend on $n$, $\alpha$, and $D_1$, the normalized recovery fraction
\[
f_U(\tau) \equiv \frac{U(T_1+\tau)-U(T_1)}{nS(D_1)-U(T_1)} = 1 - e^{-\beta\tau}
\]
cancels all $(n,\alpha,D_1)$-dependent prefactors cleanly because each damaged target in the residual injury pool $n S(D_1) - U(t)$ repairs independently at first-order rate $\beta$. Consequently, the 95\% recovery time $\tau_{95\%} = (\ln 20)/\beta$ is a universal constant independent of $n$, $\alpha$, and $D_1$ (yielding $502.6\text{ min} = 8.38\text{ h}$ for V79; Fig.~\ref{fig:split_dq}(b)), explaining the empirical ``$\approx 8.3\text{ h}$'' plateau noted numerically in Section~5.4 of \citet{shibata2026dynamic}.

\begin{figure}[tbp]
\centering
\includegraphics[width=0.96\textwidth]{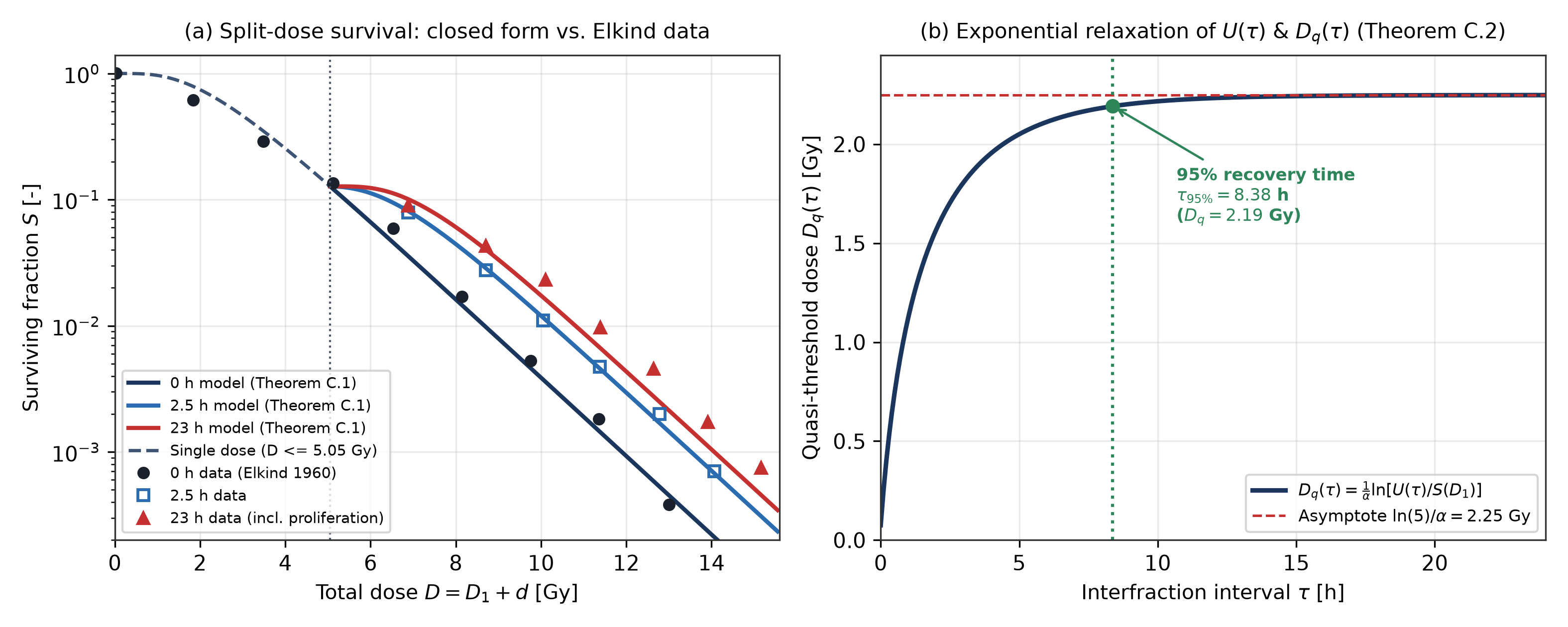}
\caption{(a) Comparison between the V79 split-dose experimental data \citep{elkind1960radiation,shibata2026dynamic} ($0\text{ h}$ black circles, $2.5\text{ h}$ open blue squares, $23\text{ h}$ red triangles) and the closed-form solution \eqref{eq:main_split} ($n=5$, $\alpha=0.716\text{ Gy}^{-1}$, $\beta=0.00596\text{ min}^{-1}$, $D_1=5.05\text{ Gy}$). (b) Recovery curve of $D_q(\tau)$ from exact identity \eqref{eq:main_Dq} and the universal 95\% recovery time $\tau_{95\%}=(\ln 20)/\beta=8.38\text{ h}$.}
\label{fig:split_dq}
\end{figure}

% =========================================================================
\section{Finding I: Chromosome-Arm Target Scale (\texorpdfstring{$\sim 100\text{ Mbp}$}{~100 Mbp}) and Genome-Size Scaling (\texorpdfstring{$G^1 \to G^n$}{G1 to Gn})}
\label{sec:finding_target_genome}
% =========================================================================

\subsection{Redundant-Hit Correction Reveals a \texorpdfstring{$\approx 90\text{--}140\text{ Mbp}$}{~90-140 Mbp} Target Size}
\label{subsec:target_arm}

Section~5.7 of \citet{shibata2026dynamic} estimated the genomic fraction occupied by critical lethal targets as $f_{\mathrm{est}} = n/(Y_{\mathrm{DSB}} M(0))$, where $Y_{\mathrm{DSB}}$ is the cellular DSB yield ($\mathrm{Gy}^{-1}$) and $M(0)$ is the acute mean inactivation dose, obtaining $3.9\text{--}7.8\%$ for $Y_{\mathrm{DSB}}=20\text{--}40\text{ Gy}^{-1}$ and contrasting the remainder with non-coding DNA. However, an inherent stochastic feature of multi-target hitting shifts this estimate substantially.

When a cell contains $n$ targets, damaging all $n$ targets at least once requires more than $n$ total hits because subsequent hits strike already-damaged targets with equal probability (redundant re-hitting). Applying Wald's identity to the stopping dose $D_{\mathrm{lethal}}$ from $i=0$ to $i=n$ (Proposition~E.1, proof in Appendix~E.1) together with the exact acute mean inactivation dose $M(0) = H_n/\alpha$ where $H_n = \sum_{k=1}^n 1/k$ is the $n$-th harmonic number (Theorem~B.2) yields the expected total number of hits before lethality:
\begin{equation}
\mathbb{E}\bigl[N_{\mathrm{hits}}^{\mathrm{total}}(0)\bigr] = n\alpha M(0) = n H_n.
\label{eq:main_wald}
\end{equation}
For V79 cells ($n=5$), $H_5 = 137/60 \approx 2.2833$, so over $M(0) = 3.189\text{ Gy}$ a cell sustains an average of \textbf{$5H_5 = 11.42$ total hits} ($5$ first-time effective hits plus $6.42$ redundant hits, hit efficiency $1/H_5 = 43.80\%$; Fig.~\ref{fig:target_genome}(a)). Thus $n/(Y_{\mathrm{DSB}} M(0))$ implicitly compressed the target pool by $1/H_5 \approx 0.438$.

Removing the redundant-hit bias $H_n$ uniquely determines both the net sensitive genomic fraction $f_{\mathrm{sens}}$ and the physical DNA length $L_1$ per target directly from $\alpha$ alone, independent of dose rate or numerical integration of $M$:
\begin{equation}
f_{\mathrm{sens}} = H_n\cdot\frac{n}{Y_{\mathrm{DSB}}M(0)} = \frac{n\alpha}{Y_{\mathrm{DSB}}}, \qquad
L_1 = \frac{\alpha}{\rho_{\mathrm{DSB}}},
\label{eq:main_L1}
\end{equation}
where $\rho_{\mathrm{DSB}} \equiv Y_{\mathrm{DSB}}/L_{\mathrm{tot}}$ is the DSB induction density per unit DNA length. Using $Y_{\mathrm{DSB}} = 20\text{--}40\text{ Gy}^{-1}$ gives a net sensitive pool of \textbf{$f_{\mathrm{sens}} = 3.58/(20\text{--}40) = 9.0\%\text{--}17.9\%$} ($1.8\%\text{--}3.6\%$ per target). Substituting standard low-LET mammalian DSB densities $\rho_{\mathrm{DSB}} = 5\text{--}8\text{ DSB/Gy/Gbp}$ \citep{ward1988dna,sachs1992dna} into Eq.~\eqref{eq:main_L1} yields
\[
L_1 = \frac{0.716\text{ Gy}^{-1}}{5\text{--}8\text{ Gy}^{-1}\text{Gbp}^{-1}} = 89.5\text{--}143.2\text{ Mbp} \quad (\approx 90\text{--}140\text{ Mbp}).
\]
The diploid Chinese hamster V79 genome has $L_{\mathrm{tot}} \approx 4.8\text{ Gbp}$ across $2n=22$ chromosomes, giving $\approx 218\text{ Mbp}$ per chromosome or \textbf{$\approx 109\text{ Mbp}$ per chromosome arm}. This quantitative match suggests that the functional ``target'' of radiation cell killing corresponds not to isolated protein-coding exons ($1\text{--}2\%$ of the genome), but rather to \textbf{chromosome-arm or large chromatin-domain integrity regions ($\sim 100\text{ Mbp}$ scale)} spanning $\sim 10\text{--}18\%$ of the genome.

\subsection{Amplification from HDR Sparrow Scaling (\texorpdfstring{$G^1$}{G1}) to LDR \texorpdfstring{$G^n$}{Gn} Scaling and WAM Universality}
\label{subsec:genome_scaling}

Once $\alpha = \rho_{\mathrm{DSB}}L_1$ links damage sensitivity directly to DNA target size, we can predict how changes in genome size $G$ across species or aneuploid variants shift radiosensitivity at arbitrary dose rates. Evaluating the terminal survival slope $\mu_1(\gamma) = 1/D_0(\gamma)$ (smallest eigenvalue of $A(b)$) via Cauchy eigenvalue interlacing on the non-absorbing Kronecker-sum matrix and determinant invariance $\det A(b) = n!\,\alpha^n$ (Theorem~B.1, Appendix~B) establishes exact two-sided bounds for all $\gamma > 0$:
\begin{equation}
\alpha\,q(\gamma)^{n-1} < \mu_1(\gamma) < n\alpha\,q(\gamma)^{n-1}, \qquad q(\gamma) \equiv \frac{\gamma}{\gamma + \gamma_*}
\label{eq:main_bounds}
\end{equation}
(illustrated across 6 decades of dose rate in Appendix~B Fig.~B.1).

\begin{figure}[tbp]
\centering
\includegraphics[width=0.96\textwidth]{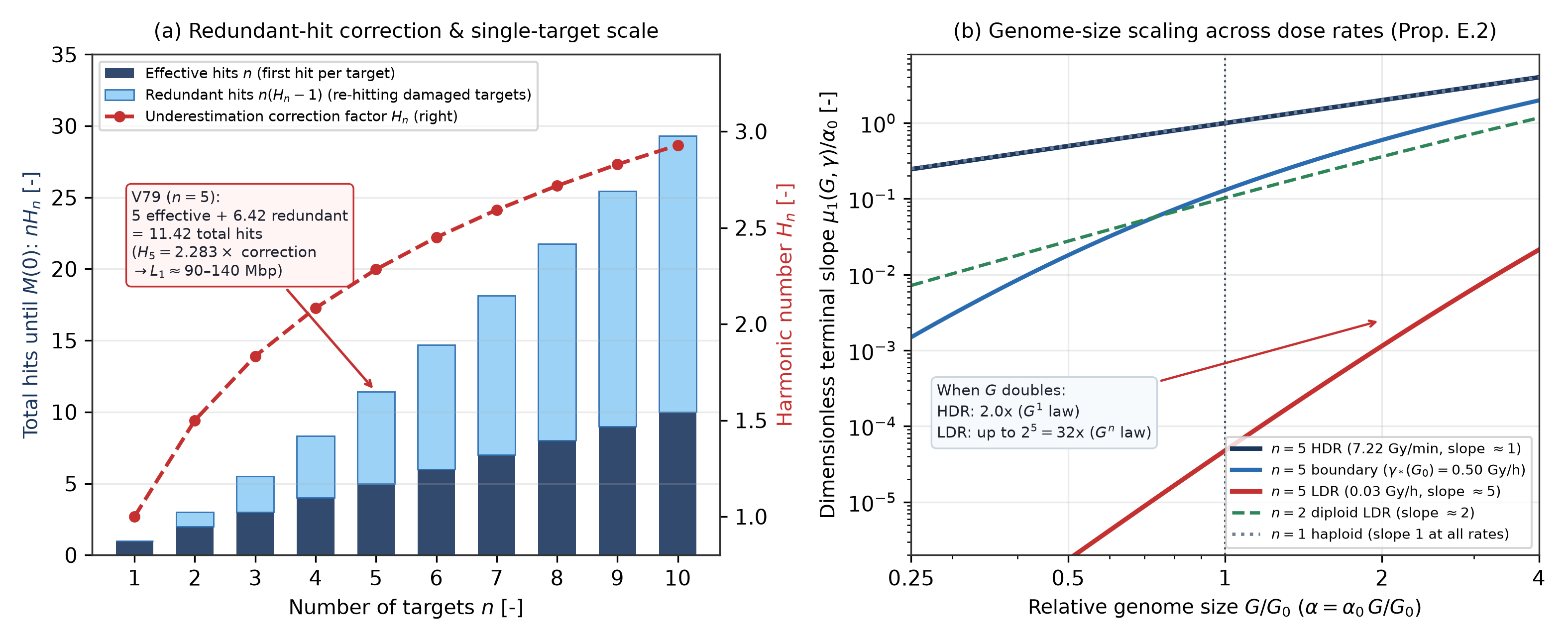}
\caption{(a) Redundant-hit correction (Proposition~E.1). At $n=5$ (V79), lethality over $M(0)$ requires $5$ effective first hits plus $6.42$ redundant hits ($11.42$ total hits, $H_5=2.283\times$), yielding $L_1 \approx 90\text{--}140\text{ Mbp}$ ($\sim 109\text{ Mbp}$ chromosome arm). (b) Genome-size scaling across dose rates (Proposition~E.2): $G^1$ Sparrow scaling at HDR vs.\ amplified $G^n$ scaling at LDR.}
\label{fig:target_genome}
\end{figure}

When cells share architecture ($n$, $\beta$) while their genome size scales by $\tilde{G} = G/G_0$, we have $\alpha(G) = \alpha_0\tilde{G}$ and $\gamma_*(G) = \gamma_*(G_0)/\tilde{G}$ (Proposition~E.2, Fig.~\ref{fig:target_genome}(b)). Consequently, the radiosensitivity ratio $\mu_1(G_A,\gamma)/\mu_1(G_B,\gamma)$ undergoes a dramatic transition across dose rates:
\begin{itemize}
  \item \textbf{HDR regime ($\gamma \gg \gamma_*$): $G^1$ scaling.} Recovery during irradiation is negligible, so $\mu_1 \approx \alpha \propto G^1$, reproducing the classical \textbf{Sparrow law} \citep{sparrow1967chromosomes} of linear scaling with nuclear DNA content or chromosome volume.
  \item \textbf{LDR regime ($\gamma \ll \gamma_*$): amplification to $G^n$ scaling.} Under competing repair, lethality requires all $n$ targets to remain damaged simultaneously with probability product $\propto q^n \sim (\alpha\gamma/\beta)^n$, yielding $\mu_1 \sim n\alpha^n(\gamma/\beta)^{n-1} \propto G^n$.
\end{itemize}
For V79 parameters ($n=5$), doubling $G$ increases the terminal slope by only $2.005\times$ at HDR ($7.22\text{ Gy/min}$), but by $4.58\times$ at $\gamma_*(G_0)\approx 0.50\text{ Gy/h}$, $23.89\times$ at $0.03\text{ Gy/h}$, and $30.07\times$ at $6\text{ mGy/h}$ (approaching $2^5 = 32\times$). By contrast, changing \emph{ploidy} $n$ at fixed per-target $\alpha$ distinguishes haploids ($n=1$, no recovery transitions $R=0$, hence no shoulder and no dose-rate effect: $\mu_1 = \alpha$) from diploids ($n=2$, where both shoulder $D_q = (\ln 2)/\alpha$ and dose-rate sparing emerge cleanly). Our framework thus separates genome-length amplification ($G^n$) from ploidy multiplicity ($n$).

\begin{figure}[tbp]
\centering
\includegraphics[width=0.96\textwidth]{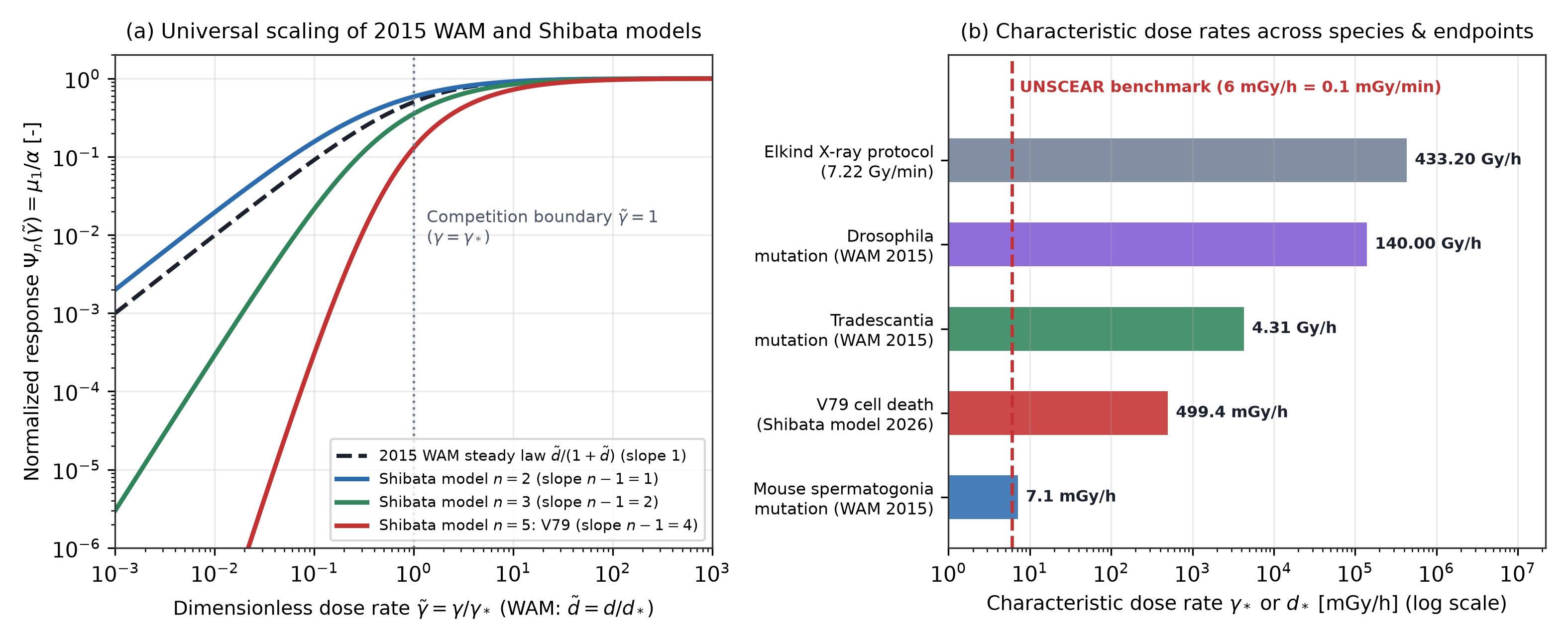}
\caption{(a) Unification of the 2015 WAM saturation law $\tilde{d}/(1+\tilde{d})$ (black dashed) and universal Shibata scaling curves $\Psi_n(\tilde{\gamma})=\mu_1(\gamma)/\alpha$ ($n=2,3,5$). The LDR log-log slope $n-1=4$ offers an independent LDR test of target number $n=5$. (b) Characteristic dose rates from WAM \citep[Table~I]{manabe2015whack} and V79 cell death \citep{shibata2026dynamic} relative to the UNSCEAR $6\text{ mGy/h}$ benchmark (red dashed).}
\label{fig:wam_scaling}
\end{figure}

Normalizing dose rate by $\tilde{\gamma} = \gamma/\gamma_*$ and slope by $\Psi_n(\tilde{\gamma}) \equiv \mu_1(\gamma)/\alpha$ eliminates species-specific $(\alpha,\beta)$ parameters and collapses cellular and organismal dose-rate responses onto a single \textbf{universal scaling family} (Fig.~\ref{fig:wam_scaling}(a)). At $n=2$, the LDR power-law slope equals $1$, matching the WAM saturation law $\tilde{d}/(1+\tilde{d})$ of \citet{manabe2015whack}, and extends to slope $n-1$ for general $n$ ($n-1=4$ for V79). Compared with UNSCEAR's empirical LDR benchmark of $6\text{ mGy/h}$ ($0.1\text{ mGy/min}$ \citep{unscear2015report}; Fig.~\ref{fig:wam_scaling}(b)), mouse spermatogonial stem-cell mutagenesis ($d_* = 7.1\text{ mGy/h}$) sits right at the benchmark, whereas V79 cell killing ($\gamma_* \approx 499\text{ mGy/h}$), \textit{Tradescantia} ($4.31\text{ Gy/h}$), and \textit{Drosophila} ($140\text{ Gy/h}$) span 2--4 orders of magnitude higher, highlighting $\gamma/\gamma_*$ as the natural dimensionless criterion.

% =========================================================================
\section{Finding II: Apparent Biphasic Recovery from a Single Repair Rate \texorpdfstring{$\beta$}{beta} and Diagnostic Tests}
\label{sec:finding_biphasic}
% =========================================================================

When split-dose shoulder restoration $D_q(\tau)$ is plotted semi-logarithmically against interfraction interval $\tau$, experiments frequently exhibit a rapid rise over $1\text{--}2\text{ h}$ followed by a slow plateau. This concave-up residual curve is routinely interpreted as a weighted sum of fast NHEJ ($\sim$tens of minutes) and slow HR ($\sim$hours) components; indeed, Sections~5.4--5.6 of \citet{shibata2026dynamic} discussed the fitted half-time of $116\text{ min}$ as a population average of fast and slow processes.

Analyzing Eq.~\eqref{eq:main_Dq} reveals a striking mathematical property independent of whether dual pathways coexist in cells: \textbf{even when true molecular repair is strictly governed by a single rate constant $\beta$, logarithmic concavity mathematically forces the macroscopic observable $D_q(\tau)$ to appear biphasic} (Proposition~E.3, proof in Appendix~E.3).

Defining $\Lambda(D_1) \equiv S(D_1)/e^{-\alpha D_1} = \sum_{k=0}^{n-1}(1-e^{-\alpha D_1})^k \in (1,n)$, the linear target recovery fraction $f_U(\tau) = 1 - e^{-\beta\tau}$ maps into the macroscopic $D_q(\tau)$ recovery fraction $f_{D_q}(\tau) \equiv [D_q(\tau)-D_q(0)]/[D_q(\infty)-D_q(0)]$ via:
\begin{equation}
f_{D_q}(\tau) = \frac{\ln\bigl[1 + (\Lambda-1)(1-e^{-\beta\tau})\bigr]}{\ln\Lambda} > f_U(\tau),
\qquad
1 - f_{D_q}(\tau) = \sum_{k=1}^{\infty} \frac{(1-\Lambda^{-1})^k}{k\ln\Lambda}\,e^{-k\beta\tau}.
\label{eq:main_biphasic}
\end{equation}
Because $\ln(\cdot)$ is strictly concave, when linear repair $f_U$ has progressed only $30.1\%$ ($\tau=1.0\text{ h}$) or $59.1\%$ ($\tau=2.5\text{ h}$), $f_{D_q}$ has already reached $48.4\%$ and $75.0\%$, compressing the apparent half-time from the true $116.3\text{ min}$ to \textbf{$63.4\text{ min}$} ($45\%$ shorter). Furthermore, expanding the residual in Eq.~\eqref{eq:main_biphasic} expresses $1 - f_{D_q}(\tau)$ strictly as a positive superposition of the true $\beta$ mode (weight $50.7\%$ for V79 at $D_1=5.05\text{ Gy}$) and faster harmonics $2\beta, 3\beta, \dots$ (combined weight $49.3\%$), accelerating the initial decay rate by $\kappa_{\mathrm{init}} = (\Lambda-1)/\ln\Lambda = \mathbf{2.403\times}$ (initial half-time $48.4\text{ min}$, further reinforced at the state level by the single-step transition rate $4\beta$ from $i=4$, half-time $\ln 2/(4\beta)=29.1\text{ min}$).

\begin{figure}[tbp]
\centering
\includegraphics[width=0.96\textwidth]{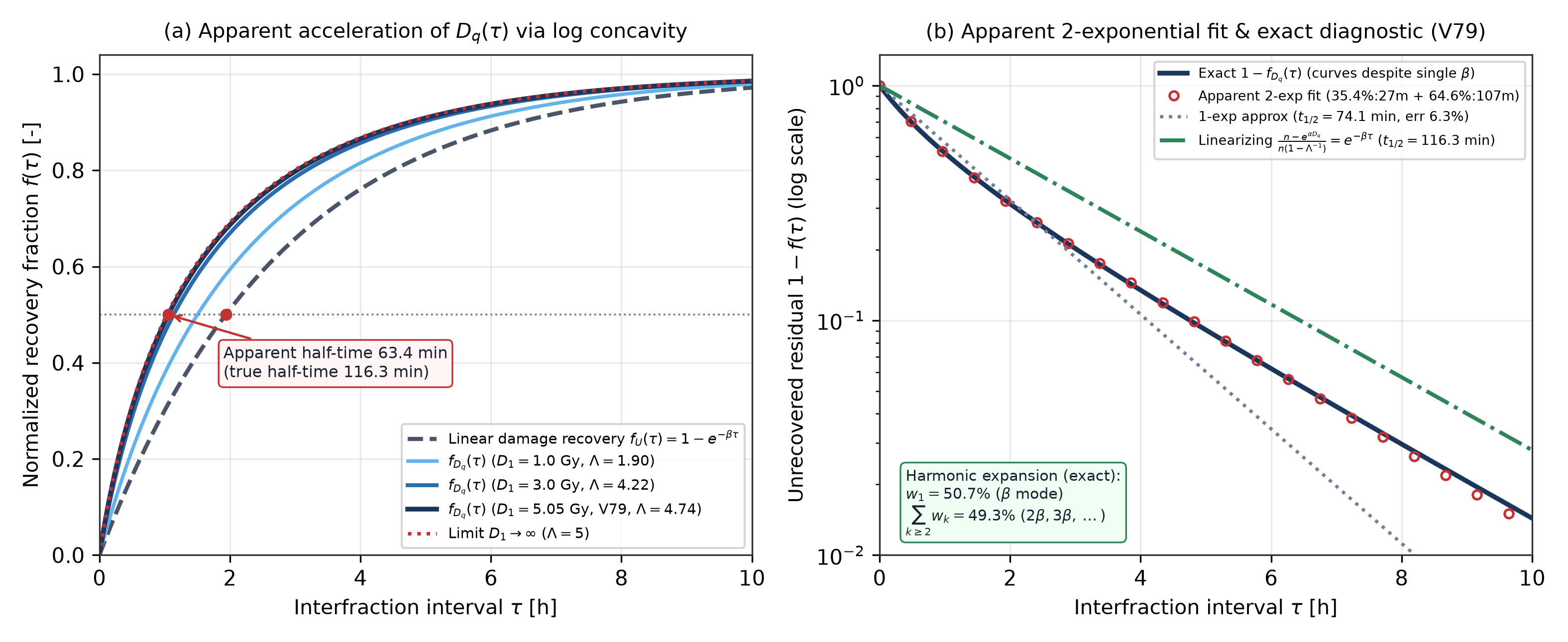}
\caption{Apparent biphasic recovery generated from a single repair rate $\beta$ via logarithmic concavity (Proposition~E.3). (a) Linear target repair $f_U(\tau)=1-e^{-\beta\tau}$ (dashed, $t_{1/2}=116.3\text{ min}$) vs.\ macroscopic shoulder recovery $f_{D_q}(\tau)$, showing apparent shortening to $63.4\text{ min}$ at $D_1=5.05\text{ Gy}$. (b) Semi-log residual $1-f_{D_q}(\tau)$: the exact single-$\beta$ curve (dark navy) bends due to harmonic terms ($50.7\%\;\beta + 49.3\%\;\ge 2\beta$) and is matched to $0.55\%$ error by a spurious two-exponential fit ($27.1\text{ min}$ [$35.4\%$] + $106.5\text{ min}$ [$64.6\%$], open red circles). Only the linearizing transform (green dash-dot) recovers the true $116.3\text{ min}$ half-time.}
\label{fig:biphasic}
\end{figure}

The practical implication for data fitting is immediate: fitting a standard bi-exponential model $w e^{-\beta_1\tau} + (1-w)e^{-\beta_2\tau}$ to the exact \emph{single-$\beta$} residual curve $1 - f_{D_q}(\tau)$ ($\tau\in[0,12]\text{ h}$) yields an apparent fast/slow decomposition of \textbf{$35.4\%$ with $t_{1/2}=27.1\text{ min}$ plus $64.6\%$ with $t_{1/2}=106.5\text{ min}$} to within $0.55\%$ maximum error, whereas a naive single-exponential fit leaves a $6.34\%$ systematic residual (Fig.~\ref{fig:biphasic}).

This finding does not deny the biological reality of NHEJ and HR; rather, it provides \textbf{two exact diagnostics} to distinguish genuine two-pathway repair from logarithmic distortion in experimental datasets:
\begin{enumerate}
  \item \textbf{Linearizing transform:} Inverting Eq.~\eqref{eq:main_Dq} gives $\ln\bigl(n - e^{\alpha D_q(\tau)}\bigr) = \ln[n(1-\Lambda^{-1})] - \beta\tau$. Under true single-$\beta$ kinetics, this transformed plot is strictly linear over all $\tau$ with unbiased slope $-\beta$ (green dash-dot line, Fig.~\ref{fig:biphasic}(b)).
  \item \textbf{Priming-dose ($D_1$) test:} Logarithmic initial acceleration $\kappa_{\mathrm{init}}(D_1) = (\Lambda(D_1)-1)/\ln\Lambda(D_1)$ grows monotonically with $D_1$ (returning to $1$ as $D_1 \to 0$), whereas a fixed physical two-pathway mechanism exhibits a $D_1$-independent fast time constant.
\end{enumerate}

% =========================================================================
\section{Finding III: A Unified Multiscale Skeleton from DNA Breaks (\texorpdfstring{$10\text{ bp}$}{10 bp}) to Stem Cells (\texorpdfstring{$42\text{ days}$}{42 days})}
\label{sec:finding_multiscale}
% =========================================================================

\subsection{Why DSB Induction Is Linear (One-Body) While Cell Killing Is Multi-Hit and Dose-Rate-Dependent}

In the low-dose limit ($\alpha D \ll 1$), the multi-stage recovery matrix $A(b)=\alpha F + bR$ yields the exact asymptotic survival law $-\ln S(D) \approx (\alpha D)^n G_n(\beta T)$ (Theorem~C.3; details and Fig.~C.1 in Appendix~C), where $G_n(x) = \frac{n}{x^n}\int_0^x (1-e^{-u})^{n-1}\,du$ generalizes the classical \textbf{Lea--Catcheside time factor} \citep{lea1942mechanism,sachs1997link} to arbitrary $n$, \textbf{reducing identically to $G_2(x) = \frac{2}{x^2}(x - 1 + e^{-x})$ at $n=2$}. For protracted exposures ($\beta T \gg 1$), $G_n(x) \sim n/x^{n-1}$, removing the apparent ``zero initial slope'' beyond $D \sim \gamma/\beta$ and establishing the linear killing rate $n\alpha^n(\gamma/\beta)^{n-1}D$.

Applying this exact $n=2$ module at the molecular scale directly models DNA single-strand break (SSB) to DSB conversion (Proposition~E.4, proof in Appendix~E.4). When two independent tracks deposit opposite-strand SSBs within interaction distance $|x_+ - x_-| \le L_c$ ($L_c \approx 10\text{--}20\text{ bp}$ \citep{ward1988dna,sachs1992dna}) before repair at rate $\beta_{\mathrm{SSB}}$, the expected two-track DSB yield is
\begin{equation}
\langle N_{\mathrm{DSB}}^{\mathrm{2\text{-}track}}(D,T)\rangle = \frac{Y_{\mathrm{SSB}}^2 L_c}{2 L_{\mathrm{tot}}}\,D^2\,G_2(\beta_{\mathrm{SSB}}T).
\label{eq:main_ssb}
\end{equation}
Equating Eq.~\eqref{eq:main_ssb} with one-track direct DSB induction ($Y_{\mathrm{DSB}}D = \rho_{\mathrm{DSB}}L_{\mathrm{tot}}D$) cancels total genome length $L_{\mathrm{tot}}$ identically, yielding a \textbf{genome-length-invariant local crossover dose}:
\begin{equation}
D_{\mathrm{cross}}(T) = \frac{2\rho_{\mathrm{DSB}}}{\rho_{\mathrm{SSB}}^2 L_c\,G_2(\beta_{\mathrm{SSB}}T)}.
\label{eq:main_dcross}
\end{equation}
For typical mammalian densities ($\rho_{\mathrm{SSB}} = 200$~SSB/Gy/Gbp, $\rho_{\mathrm{DSB}} = 7$~DSB/Gy/Gbp, $L_c = 10\text{--}20\text{ bp}$), the acute crossover dose is \textbf{$D_{\mathrm{cross}}(0) = 17.5\text{--}35.0\text{ kGy}$}. At $D = 10\text{ Gy}$, two-track SSB coincidence contributes only $0.10\text{--}0.20$ DSBs per cell ($0.03\text{--}0.06\%$ of the $350$ direct DSBs) and is suppressed even further under protracted irradiation by $G_2(\beta_{\mathrm{SSB}}T)$ (Fig.~\ref{fig:multiscale}(a)).

Therefore, the reason DSB induction appears linear at multi-Gy doses whereas cell killing exhibits strong multi-hit dose-rate dependence is not a change in mathematical law, but a \textbf{$10^7$-fold expansion of target length from $L_c \sim 10\text{ bp}$ ($\mathrm{SSB}\to\mathrm{DSB}$) to $L_1 \sim 100\text{ Mbp}$ (V79 chromosome arm)}, which amplifies per-unit radiosensitivity from $\sim 10^{-6}\text{ Gy}^{-1}$ to $0.716\text{ Gy}^{-1}$ ($\sim 10^6$-fold).

\subsection{Unified Space--Time Hierarchy Map Across Molecules, Cells, and Tissues}

\begin{figure}[tbp]
\centering
\includegraphics[width=0.96\textwidth]{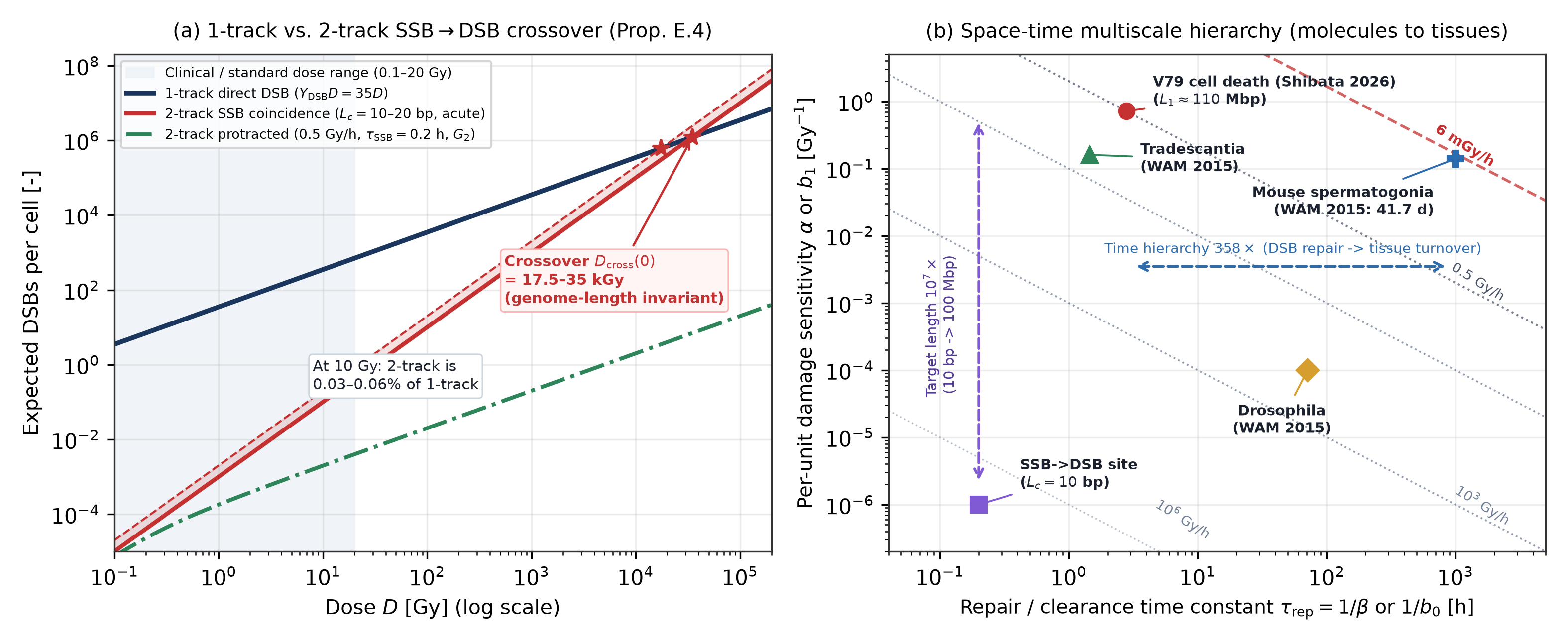}
\caption{(a) Dose dependence of expected cellular DSBs (Proposition~E.4). One-track direct DSBs dominate across clinical and experimental doses ($0.1\text{--}20\text{ Gy}$), while two-track SSB coincidence ($L_c=10\text{--}20\text{ bp}$) overtakes only above the genome-length-invariant $D_{\mathrm{cross}}(0)=17.5\text{--}35\text{ kGy}$. (b) Multiscale space--time hierarchy map of $\tau_{\mathrm{rep}}$ vs.\ $\alpha$ (or $b_1$; red dashed line: UNSCEAR $6\text{ mGy/h}$), cleanly separating the $10^7$-fold spatial jump ($\mathrm{SSB}\to\text{DSB}$ to cell death) from the $358$-fold temporal jump (cellular repair to stem-cell turnover).}
\label{fig:multiscale}
\end{figure}

Plotting repair/clearance time constant $\tau_{\mathrm{rep}} = 1/\beta$ (or $1/b_0$) horizontally against per-unit damage sensitivity $\alpha$ (or $b_1$) vertically organizes all three biological tiers onto a single phase plane governed by the same one-body and multi-stage recovery-competition modules (Fig.~\ref{fig:multiscale}(b), Table~\ref{tab:hierarchy}). Vertically, $\mathrm{SSB}\to\mathrm{DSB}$ coincidence ($L_c=10\text{ bp}$, $\tau_{\mathrm{rep}}\sim 0.2\text{ h}$) and V79 cell killing ($L_1\sim 100\text{ Mbp}$, $\tau_{\mathrm{rep}}=2.80\text{ h}$, $\gamma_*\approx 0.50\text{ Gy/h}$) are separated by the $\sim 10^7$-fold spatial target ratio. Horizontally, V79 cell killing and mouse spermatogonial stem-cell mutagenesis from \citet{manabe2015whack} ($\tau_{\mathrm{rep}} = 1/b_0 = 1000\text{ h} = 41.7\text{ d}$, $d_* = 7.1\text{ mGy/h}$) are separated by a $358$-fold temporal hierarchy bridging rapid cellular DSB repair and slow tissue stem-cell turnover.

\begin{table}[tbp]
\centering
\caption{Two recurring mathematical modules across molecular, cellular, and tissue-level low-LET radiation responses.}
\label{tab:hierarchy}
\footnotesize
\begin{tabularx}{\textwidth}{@{}>{\raggedright\arraybackslash}p{2.9cm}>{\raggedright\arraybackslash}p{4.8cm}>{\raggedright\arraybackslash}X@{}}
\toprule
Biological Tier & Single-track induction\newline (dose-rate-independent total yield) & Multi-stage / multi-target recovery competition\newline ($A(b)$ and $G_n(\beta T)$) \\
\midrule
$\mathrm{SSB} \to \mathrm{DSB}$ & Single-track clustered lesion $\implies$ linear $Y_{\mathrm{DSB}}D$ & Independent two-track opposite-strand SSBs ($|x_+-x_-|\le L_c\sim 10\text{ bp}$) competing with $\beta_{\mathrm{SSB}}$ $\implies n=2$: $\frac{Y_{\mathrm{SSB}}^2 L_c}{2L_{\mathrm{tot}}}D^2 G_2(\beta_{\mathrm{SSB}}T)$ \\
DSB $\to$ Mutation / Aberration & Single-DSB misrepair (Path 1: $1-e^{-C_1 D}$, \citet{kingman1993poisson}) & Pairwise DSB exchange (Path 2: $\propto D^2 G_2(\beta_{s2}T)$, \citet{sachs1997link}) and WAM steady balance \citep{manabe2015whack} \\
Lethal lesion $\to$ Cell death & Irreparable single-hit lethal lesion ($+\alpha_0 I \implies e^{-\alpha_0 D}$) & Damage accumulation across $n$ chromosome-arm targets ($\sim 100\text{ Mbp}$) $\implies \mathbf{1}^T e^{-A(b)D}\mathbf{p}_0$, reducing to $(\alpha D)^n G_n(\beta T)$ at low dose \\
\bottomrule
\end{tabularx}
\end{table}

% =========================================================================
\section{Dip Mechanism, Post-Recovery Proliferation, and Experimental Tests}
\label{sec:implications}
% =========================================================================

\subsection{Cell-Cycle Sensitivity \texorpdfstring{$\alpha(\tau)$}{alpha(tau)}, Not Recovery Rate \texorpdfstring{$\beta(t)$}{beta(t)}, Drives the Survival Dip}

In \citet{elkind1960radiation} Fig.~16 (two-fraction survival over $\tau \in [0,12]\text{ h}$), survival transiently drops around $\tau = 4\text{--}6\text{ h}$ before rising again (the ``Elkind dip''). Section~4.6 and Fig.~7 of \citet{shibata2026dynamic} compared a time-varying recovery rate $\beta(t)$ against a time-varying sensitivity $\alpha(\tau)$.

From first principles, even if an instantaneous nonnegative repair rate $\beta(t) \ge 0$ dips transiently, repair cannot run backward to create new lesions; hence the cumulative recovery integral $\Phi_{\mathrm{int}}(\tau) \equiv \int_0^\tau \beta(t)\,dt$ is nondecreasing in $\tau$. Formalizing this intuition proves rigorously (Theorem~D.1, Appendix~D) that \textbf{whenever $d\mathbf{p}/dt = -\beta(t)R\mathbf{p}$ is integrated consistently with any $\beta(t) \ge 0$, post-second-dose survival $S(\tau)$ is strictly monotonically increasing in $\Phi_{\mathrm{int}}(\tau)$ and can never produce a dip.}

\begin{figure}[tbp]
\centering
\includegraphics[width=0.96\textwidth]{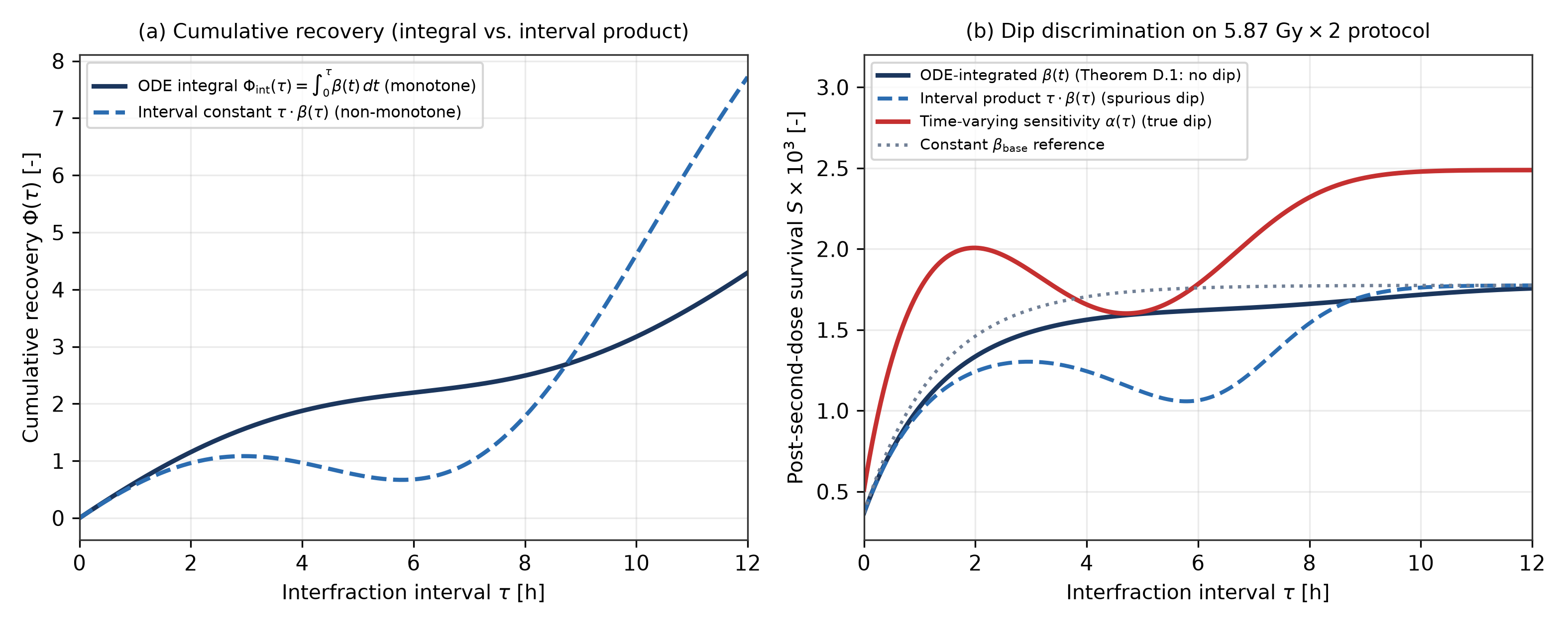}
\caption{Mechanism discrimination for the $5.87\text{ Gy}\times 2$ survival dip using Table~3 parameters of \citet{shibata2026dynamic} (Theorem~D.1). (a) Cumulative recovery: exact ODE integration $\Phi_{\mathrm{int}}(\tau)=\int_0^\tau\beta(t)\,dt$ is strictly nondecreasing, whereas the interval product $\tau\beta(\tau)$ drops over $3\text{--}6\text{ h}$. (b) Post-second-dose survival: ODE-integrated $\beta(t)$ (dark navy) produces no dip; time-dependent sensitivity $\alpha(\tau)$ (red) generates the genuine cell-cycle dip.}
\label{fig:dip}
\end{figure}

Evaluating Table~3 of \citet{shibata2026dynamic} directly confirms this theorem (Fig.~\ref{fig:dip}): the true ODE time integral $\Phi_{\mathrm{int}}(\tau)$ yields a strictly monotonic survival curve (dark navy solid line). The dip shown in Fig.~7(a) of \citet{shibata2026dynamic} (dashed blue curve) resulted from evaluating interval-wise constant products $\tau\cdot\beta(\tau)$, whereas consistent ODE evolution generates a genuine dip only through time-varying sensitivity $\alpha(\tau)$ reflecting cell-cycle redistribution into radiosensitive G2/M phases (red curve). Biologically, combining primary $\alpha(\tau)$ modulation with consistent ODE-integrated $\beta(t)$ within the same matrix framework is straightforward.

\subsection{Modular Extensions: Single-Hit Killing and G2-Checkpoint Proliferation with Colony Multiplicity}

\begin{figure}[tbp]
\centering
\includegraphics[width=0.96\textwidth]{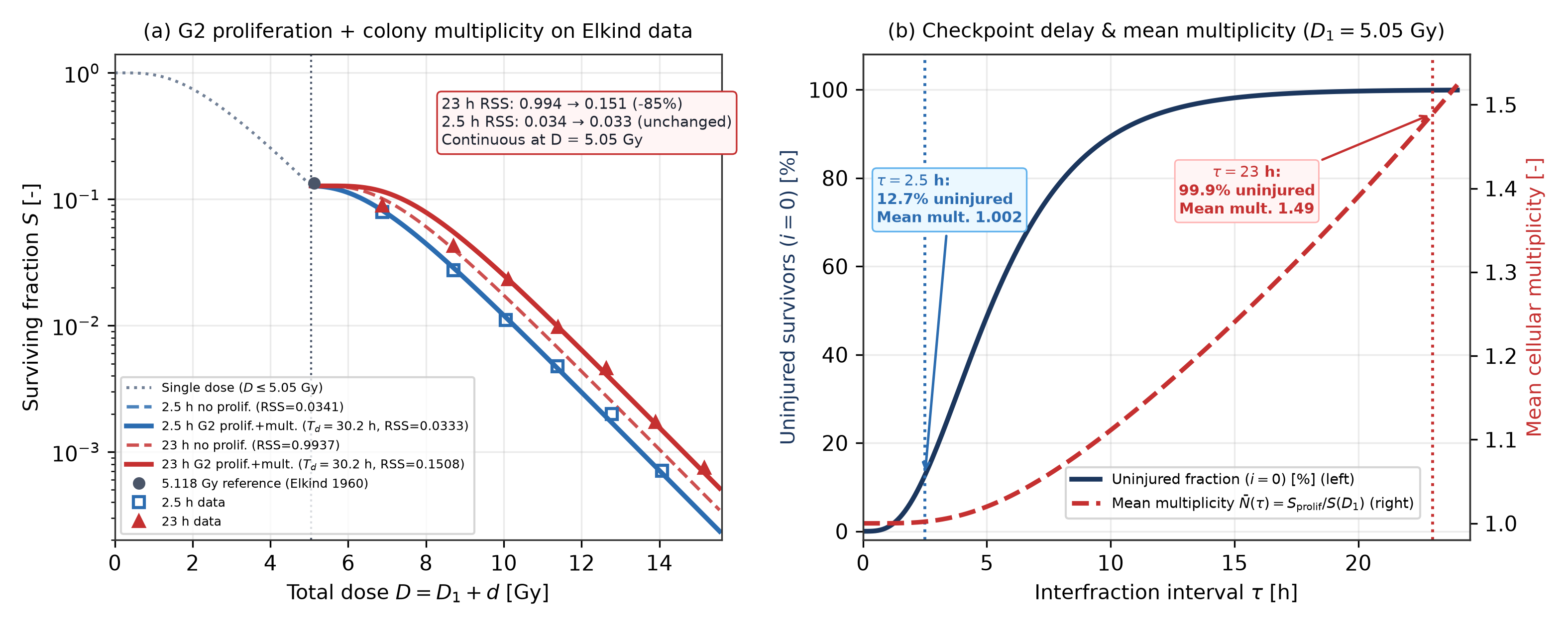}
\caption{Verification of G2-checkpoint-gated proliferation via the 1-entry matrix extension $\beta R - G$ coupled with microcolony multiplicity. (a) Comparison against Elkind--Sutton V79 data ($2.5\text{ h}$ open blue squares, $23\text{ h}$ red triangles) without proliferation (dashed) and with G2 proliferation + multiplicity (solid, $T_d=30.2\text{ h}$, $\bar{N}_{23\mathrm{h}}=1.49$). Continuity with the single-dose curve at $D=5.05\text{ Gy}$ and the $2.5\text{ h}$ fit are preserved while cutting $23\text{ h}$ $\mathrm{RSS}$ by $85\%$. (b) Uninjured survivor ($i=0$) percentage (left axis, navy solid) and mean cellular multiplicity $\bar{N}(\tau)$ (right axis, red dashed) after $D_1=5.05\text{ Gy}$.}
\label{fig:g2_prolif}
\end{figure}

The matrix formulation~\eqref{eq:main_matrix} accommodates additional radiobiological effects by simple single-term modifications without rebuilding equations from scratch:
\begin{enumerate}
  \item \textbf{Irreparable single-hit killing:} Adding a single-hit lethal rate $\alpha_0$ ($\mathrm{Gy}^{-1}$) from every surviving state shifts $A_{\mathrm{ext}}(b) = A(b) + \alpha_0 I$, factoring survival strictly as $S_{\mathrm{ext}}(D) = e^{-\alpha_0 D}S(D)$ and setting a nonzero LDR limiting slope $\alpha_0 > 0$.
  \item \textbf{Post-recovery proliferation (G2-checkpoint arrest and colony multiplicity):} The underestimation of $23\text{ h}$ split-dose survival noted in \citet{shibata2026dynamic} arises from cell division after repair completes. Even the realistic mechanism in which cells remain arrested at the G2 DNA-damage checkpoint while damaged ($i \ge 1$) and proliferate at rate $g$ only after full recovery to $i=0$ is encoded simply by changing the single top-left entry $(0,0)$ of $\beta R$ from $0$ to $-g$, forming $\beta R - G$ (Appendix~A.1). Here the net cellular multiplication factor $\bar{N}(\tau) \equiv \mathbf{1}^T e^{-(\beta R-G)\tau}\mathbf{p}(T_1)/S(D_1)$ directly predicts single-cell suspension re-plating assays (where cell count multiplies linearly by $\bar{N}(\tau)$, creating a step at $D_1$). In \citet{elkind1960radiation}'s attached Petri-dish colony assay, however, daughter cells remain within the same microcolony of mean cellular multiplicity $\bar{N}(\tau)$; colony survival obeys the classical multiplicity relation $S_{\mathrm{colony}}(D_1,\tau,d) = S(D_1)\{1 - [1 - S_{\mathrm{rel}}(d)]^{\bar{N}(\tau)}\}$ (with $S_{\mathrm{rel}}(d)$ denoting conditional single-cell survival to $d$), remaining continuous with $S(D_1)$ at $d=0$ ($D=5.05\text{ Gy}$) while approaching $\bar{N}(\tau)S(D_1)S_{\mathrm{rel}}(d)$ at high doses. Applying this formulation with $(n,\alpha,\beta)$ held fixed at Shibata's values keeps the $i=0$ fraction at only $12.7\%$ ($\bar{N}=1.002$) at $\tau=2.5\text{ h}$, leaving $2.5\text{ h}$ $\mathrm{RSS}$ unchanged ($0.0341 \to 0.0333$), while at $\tau=23\text{ h}$ $99.9\%$ of survivors return to $i=0$ ($\bar{N}=1.49$, effective doubling time $T_d=30.2\text{ h}$), cutting $23\text{ h}$ $\mathrm{RSS}$ by $\sim 85\%$ from $0.9937$ to $0.1508$ with exact continuity at $D=5.05\text{ Gy}$ (Fig.~\ref{fig:g2_prolif}). As a reference check, jointly optimizing $(\alpha,\beta,T_d)$ across all $22$ Elkind data points ($0, 2.5, 23\text{ h}$) yields $\alpha=0.707\text{ Gy}^{-1}$, $\beta=0.00503\text{ min}^{-1}$, and $T_d=39.5\text{ h}$ ($\bar{N}_{23\mathrm{h}}=1.33$), barely shifting the baseline parameters while improving all three series simultaneously ($\mathrm{RSS}$: $0\text{ h}$, $0.299 \to 0.267$; $2.5\text{ h}$, $0.034 \to 0.023$; $23\text{ h}$, $0.994 \to 0.148$; total $1.327 \to 0.438$, $-67\%$).
\end{enumerate}

\subsection{Falsifiable Experimental Tests and Scope Limitations}

Every scaling law derived here yields quantitative, parameter-free predictions testable by experiment (Table~\ref{tab:falsifiable_tests}). For instance, the target number $n=5$ inferred solely from the HDR shoulder of V79 cells predicts that across proliferation-corrected LDR survival curves spanning $\gamma_* \approx 0.50\text{ Gy/h}$ (e.g., $0.05\text{--}6.0\text{ Gy/h}$), the terminal slope $\mu_1(\gamma)$ will follow a log-log slope of $n-1 = 4$ (or flatten toward an $\alpha_0 > 0$ floor if single-hit killing is present). Because quantitative comparison here relies primarily on V79 data, systematic testing across other cell lines and species is essential.

\begin{table}[tbp]
\centering
\caption{Summary of falsifiable biophysical predictions and experimental validation protocols.}
\label{tab:falsifiable_tests}
\footnotesize
\begin{tabularx}{\textwidth}{@{}>{\raggedright\arraybackslash}p{2.5cm}>{\raggedright\arraybackslash}p{4.6cm}>{\raggedright\arraybackslash}p{3.9cm}>{\raggedright\arraybackslash}X@{}}
\toprule
Prediction & Exact Relation & V79 / Mammalian Value & Experimental Test \\
\midrule
1.\ Target size\newline (Prop.~E.1) & $\mathbb{E}[N_{\mathrm{hits}}^{\mathrm{total}}(0)]=n H_n$\newline $L_1=\alpha/\rho_{\mathrm{DSB}}$ & $H_5=2.283$ ($11.42$ total hits)\newline $L_1\approx 90\text{--}140\text{ Mbp}$ (V79 arm $\sim 109\text{ Mbp}$) & Compare $\alpha/\rho_{\mathrm{DSB}}$ against chromosome-arm/domain sizes across karyotypes \\
2.\ Genome law\newline (Prop.~E.2) & $\mu_1^{\mathrm{HDR}}\propto G^1$ (Sparrow)\newline $\mu_1^{\mathrm{LDR}}\propto G^n$ & Doubling $G$ gives:\newline HDR $2.00\times$ vs.\ LDR $23.9\text{--}30.1\times$ (limit $32\times$) & Compare HDR vs.\ LDR terminal-slope ratios across related species or aneuploids with shared $n$ \\
3.\ LDR slope\newline (Thm.~B.1) & $\mu_1(\gamma)\sim n\alpha^n(\gamma/\beta)^{n-1}$\newline $\gamma_*=\beta/\alpha$ & $\gamma_*\approx 0.50\text{ Gy/h}$\newline log-log slope $n-1=4$ ($c_1:5\to 1$) & Measure $1/D_0(\gamma)$ across $\gamma_*$ to test slope $n-1$ or detect $\alpha_0>0$ floor \\
4.\ Apparent 2-phase\newline (Prop.~E.3) & $\ln(n-e^{\alpha D_q})=\text{const.}-\beta\tau$\newline $\kappa_{\mathrm{init}}(D_1)=(\Lambda-1)/\ln\Lambda$ & Single $\beta$ ($116.3\text{ min}$) mimics $27.1\text{ min}+106.5\text{ min}$ & Apply linearizing transform to $D_q(\tau)$ and measure $\kappa_{\mathrm{init}}(D_1)$ across priming doses $D_1$ \\
5.\ Multiscale\newline (Prop.~E.4) & $D_{\mathrm{cross}}(T)=\dfrac{2\rho_{\mathrm{DSB}}}{\rho_{\mathrm{SSB}}^2 L_c G_2(\beta_{\mathrm{SSB}}T)}$ & $D_{\mathrm{cross}}(0)=17.5\text{--}35\text{ kGy}$ ($L_{\mathrm{tot}}$-free)\newline $L_1/L_c\sim 10^7$, WAM ratio $358\times$ & Test nonlinear DSB yields at kGy doses and $d_*\to\gamma_*$ shift upon tissue-turnover suppression \\
\bottomrule
\end{tabularx}
\end{table}

Key limitations and future extensions include: (a) correlated multi-target transitions induced by single high-LET tracks, (b) explicit two-dimensional Markov chains separating distinct fast/slow repair pathways (NHEJ vs.\ HR), (c) nonlinear repair saturation, (d) non-targeted tissue responses such as bystander and immune effects, and (e) explicit inclusion of non-irradiated spontaneous damage and \textbf{spontaneous / background cell death} analogous to the spontaneous mutation rate ($a_0$) in \citet{manabe2015whack}, which will become important under ultra-low-dose-rate chronic exposures and tissue-homeostasis modeling.

% =========================================================================
\section{Conclusion}
\label{sec:conclusion}
% =========================================================================

We have analytically reformulated the dynamic multi-target recovery model of \citet{shibata2026dynamic} as a natural multi-stage extension of the WAM framework \citep{manabe2015whack}, yielding a unified, falsifiable biophysical interpretation of four central questions in radiobiology. Based primarily on V79 cell data and published WAM parameters, three core insights emerge. First, correcting for redundant hits ($H_n$) brings single-target DNA size into quantitative agreement with the chromosome-arm scale ($\sim 100\text{ Mbp}$) and predicts dose-rate amplification of genome-size sensitivity from $G^1$ (Sparrow's law) to $G^n$. Second, logarithmic concavity in $D_q(\tau)=\alpha^{-1}\ln[U(\tau)/S(D_1)]$ generates apparent biphasic recovery from a single repair rate $\beta$, resolvable by our linearizing transform and $D_1$-variation test. Third, $\mathrm{SSB}\to\mathrm{DSB}$ coincidence ($n=2$), cellular lethality, and tissue stem-cell mutagenesis share a common mathematical structure separated cleanly by spatial ($\sim 10^7$-fold) and temporal ($\sim 358$-fold) scales. Within the same formulation, the split-dose dip arises from cell-cycle sensitivity modulation $\alpha(\tau)$ rather than $\beta(t)$, and post-recovery proliferation with colony multiplicity is captured via a single matrix-entry modification. Systematic application across diverse cell lines and low-dose-rate experiments is warranted to test the generality of these predictions.

\section*{Acknowledgments}
We are deeply grateful to Renta Shibata for constructing the dynamic sub-lethal damage recovery model that motivated this analytical study and for making the digitized Elkind--Sutton experimental dataset publicly available (DOI: \texttt{10.17632/3wrx25by4s.1}).

\section*{Declaration of Generative AI Use}
Interactive generative AI (Google Gemini / Antigravity) was used as an auxiliary tool in drafting numerical verification scripts, generating figures, and refining manuscript structure and English exposition. All mathematical proofs, numerical calculations, biophysical interpretations, reference verifications, and the final content of this work are the sole responsibility of the authors.

% =========================================================================
% APPENDICES: COMPLETE MATHEMATICAL FORMULATION AND PROOFS
% =========================================================================
\clearpage
\appendix
\counterwithin{equation}{section}
\counterwithin{figure}{section}
\counterwithin{table}{section}
\renewcommand{\theequation}{\Alph{section}.\arabic{equation}}
\renewcommand{\thefigure}{\Alph{section}.\arabic{figure}}
\renewcommand{\thetable}{\Alph{section}.\arabic{table}}
\renewcommand{\thetheorem}{\Alph{section}.\arabic{theorem}}
\renewcommand{\thelemma}{\Alph{section}.\arabic{lemma}}
\renewcommand{\thecorollary}{\Alph{section}.\arabic{corollary}}
\renewcommand{\theproposition}{\Alph{section}.\arabic{proposition}}

\section{Matrix Representation \texorpdfstring{$A(b)=\alpha F+bR$}{A(b)=alpha F + bR}, Nomenclature, Fundamental Theorems, and Numerical Verification}
\label{app:matrix}

\subsection{Unified Nomenclature, Matrices \texorpdfstring{$F$}{F} and \texorpdfstring{$R$}{R}, and G2-Checkpoint Proliferation Matrix}
\label{app:notation}

To avoid symbol collisions across scales (e.g., eigenvalues $\mu_k$ vs.\ damage rates, tier-specific recovery ratios $b$), Table~\ref{tab:nomenclature} summarizes the unified notation used throughout this paper alongside the original symbols of \citet{shibata2026dynamic}.

\begin{table}[htbp]
\centering
\caption{Unified notation, correspondence to \citet{shibata2026dynamic} (\textit{J.\ Theor.\ Biol.}\ 637, 112620), and V79 parameter values.}
\label{tab:nomenclature}
\footnotesize
\begin{tabularx}{\textwidth}{@{}>{\raggedright\arraybackslash}p{2.2cm}>{\raggedright\arraybackslash}p{1.6cm}>{\raggedright\arraybackslash}p{4.5cm}>{\raggedright\arraybackslash}p{2.4cm}>{\raggedright\arraybackslash}X@{}}
\toprule
Unified Symbol & Shibata (2026) & Definition & Units & V79 / Protocol Value \\
\midrule
$D,\ \gamma,\ T$ & $D,\ \gamma,\ T$ & Total dose, constant dose rate, irradiation time $T=D/\gamma$ & $\mathrm{Gy},\ \mathrm{Gy/min},\ \mathrm{min}$ & $D_1=5.05\text{ Gy},\ \gamma=7.22\text{ Gy/min},\ T_1=0.699\text{ min}$ \\
$\tau$ & $t_{\text{rec}}$, etc. & Radiation-free split-dose recovery interval & $\mathrm{min}$ or $\mathrm{h}$ & $0,\ 2.5,\ 23\text{ h}$ ($0,\ 150,\ 1380\text{ min}$) \\
$n$ & $n$ & Number of lethal targets per cell & dimensionless & $5$ (rounded from fitted $4.9$) \\
$\alpha$ & $\kappa$ & Per-target damage induction rate per unit dose & $\mathrm{Gy}^{-1}$ & $0.716\text{ Gy}^{-1}$ (rounded $0.7$) \\
$\beta$ & $\rho$ & Per-target first-order recovery rate per unit time & $\mathrm{min}^{-1}$ & $0.00596\text{ min}^{-1}$ (rounded $0.006$) \\
$b$ & $\rho/\gamma$ & Per-dose recovery rate $b=\beta/\gamma$ & $\mathrm{Gy}^{-1}$ & $8.25\times 10^{-4}\text{ Gy}^{-1}$ at $7.22\text{ Gy/min}$ \\
$\gamma_*$ & (new) & Characteristic dose rate $\gamma_*=\beta/\alpha$ ($\alpha=b$ boundary) & $\mathrm{Gy/min}$ & $0.008324\text{ Gy/min} \approx 0.50\text{ Gy/h}$ \\
$F,\ R$ & (new) & Fixed integer Forward and Recovery matrices & dimensionless & Determined solely by $n$ \\
$A,\ \bar{A},\ \tilde{A}$ & (new) & $n\times n$ surviving generator, $(n+1)\times(n+1)$ non-absorbing matrix, symmetrized $\tilde{A}=WAW^{-1}$ & $\mathrm{Gy}^{-1}$ & $A(b)=\alpha F + b R$ \\
$\mu_1 < \dots < \mu_n$ & (new) & Simple positive real eigenvalues of $A$ & $\mathrm{Gy}^{-1}$ & $\mu_k = k\alpha$ at $b=0$ \\
$S(D),\ D_0,\ c_1$ & $S(t),\ D_0,\ n$ & Surviving fraction, reciprocal terminal slope $D_0=1/\mu_1$, extrapolation number $c_1$ & $-,\ \mathrm{Gy},\ -$ & $D_0=1.397\text{ Gy},\ c_1=5$ at $b=0$ \\
$M(b)$ & $M$ & Mean inactivation dose $\int_0^\infty S(D)\,dD$ & $\mathrm{Gy}$ & $3.189\text{ Gy}$ at $b=0$ ($3.262\text{ Gy}$ for $\alpha=0.7$) \\
\bottomrule
\end{tabularx}
\end{table}

Acting on the surviving state vector $\mathbf{p}(D) = (p^{(0)}(D),\dots,p^{(n-1)}(D))^T$, the nonzero entries ($0\le i\le n-1$) of the $n\times n$ matrices $F$ and $R$ in Eq.~\eqref{eq:main_matrix} are
\begin{equation}
F_{i,i} = n-i, \quad F_{i+1,i} = -(n-i), \qquad R_{i,i} = i, \quad R_{i,i+1} = -(i+1),
\label{eq:app_FR}
\end{equation}
which for $n=5$ read
\[
F = \begin{pmatrix}
5 & 0 & 0 & 0 & 0 \\
-5 & 4 & 0 & 0 & 0 \\
0 & -4 & 3 & 0 & 0 \\
0 & 0 & -3 & 2 & 0 \\
0 & 0 & 0 & -2 & 1
\end{pmatrix}, \qquad
R = \begin{pmatrix}
0 & -1 & 0 & 0 & 0 \\
0 & 1 & -2 & 0 & 0 \\
0 & 0 & 2 & -3 & 0 \\
0 & 0 & 0 & 3 & -4 \\
0 & 0 & 0 & 0 & 4
\end{pmatrix}.
\]
In the appendices we adopt standard Markov-process terminology, referring to the irreversible lethal state $n$ as an \emph{absorbing state}, and to the $(n+1)\times(n+1)$ non-absorbing matrix allowing hypothetical recovery from state $n$ as $\bar{A}$. G2-checkpoint-gated proliferation during a radiation-free interval ($\gamma=0$) replaces only the $(0,0)$ entry of $\beta R$ by $-g$ to yield $\beta R - G$ ($G = \mathrm{diag}(g,0,\dots,0)$), giving the exact one-line solution $d\mathbf{p}/dt = -(\beta R - G)\mathbf{p}$, or $\mathbf{p}_{\mathrm{prolif}}(T_1+\tau) = \exp[-(\beta R - G)\tau]\mathbf{p}(T_1)$. The mean cellular multiplicity is then $\bar{N}(\tau) = \mathbf{1}^T\mathbf{p}_{\mathrm{prolif}}(T_1+\tau)/S(D_1)$; defining conditional single-cell survival to the second dose $d$ as $S_{\mathrm{rel}}(d) = \mathbf{1}^T e^{-\alpha F d}\mathbf{p}_{\mathrm{prolif}}(T_1+\tau)/\mathbf{1}^T\mathbf{p}_{\mathrm{prolif}}(T_1+\tau)$, attached-dish colony survival follows as $S_{\mathrm{colony}}(D_1,\tau,d) = S(D_1)\{1 - [1 - S_{\mathrm{rel}}(d)]^{\bar{N}(\tau)}\}$.

\subsection{Phase-Type Representation, Spectral Decomposition, and Determinant Invariance}
\label{app:phase_det}

\begin{theorem}[Matrix Exponential, Keilson Spectral Decomposition, and Determinant Invariance]
\label{thm:phase_type}
For any $\alpha > 0$ and $b \ge 0$:
\begin{enumerate}
  \item For initial state $\mathbf{p}_0$, $\mathbf{p}(D) = e^{-A(b)D}\mathbf{p}_0$ and $S(D) = \mathbf{1}^T e^{-A(b)D}\mathbf{p}_0$ \citep{neuts1981matrix}; arbitrary multi-fraction protocols satisfy $\mathbf{p}_{\mathrm{end}} = \bigl(\prod_m e^{-A(b_m)D_m}e^{-\beta\tau_m R}\bigr)\mathbf{p}_0$.
  \item All eigenvalues of $A(b)$ are simple positive real numbers $0 < \mu_1 < \mu_2 < \dots < \mu_n$, and for $\mathbf{p}_0 = \mathbf{e}_0$,
  \begin{equation}
  S(D) = \sum_{k=1}^{n} c_k e^{-\mu_k D}, \qquad c_k = \prod_{j\neq k}\frac{\mu_j}{\mu_j-\mu_k} \quad (c_1 > 1)
  \label{eq:keilson}
  \end{equation}
  \citep{keilson1979markov,parlett1998symmetric}. Numerical evaluation via \texttt{scipy.linalg.expm} \citep{almohy2010new} avoids cancellation.
  \item The determinant of $A(b)$ is strictly independent of $b$: $\det A(b) = \prod_{k=1}^n \mu_k = n!\,\alpha^n > 0$.
\end{enumerate}
\end{theorem}

\begin{proof}
(2) For $b>0$, similarity transformation $WAW^{-1}$ via $W = \mathrm{diag}(w_0,\dots,w_{n-1})$ with $w_{i+1} = w_i\sqrt{\frac{(i+1)b}{(n-i)\alpha}}$ produces an irreducible real symmetric tridiagonal matrix with simple real eigenvalues \citep{parlett1998symmetric} (at $b=0$, $A$ is lower triangular with eigenvalues $k\alpha$). (3) Column sums of $A(b)$ vanish for $j \le n-2$ ($-jb + [(n-j)\alpha + jb] - (n-j)\alpha = 0$) and equal $-(n-1)b + [\alpha + (n-1)b] = \alpha$ for $j=n-1$, so $\mathbf{1}^T A = (0,\dots,0,\alpha)$. Adding rows $0$ through $n-2$ to the last row and expanding along the last row yields $\det A = \alpha\det A^{(n-1)}$, where the leading $(n-1)\times(n-1)$ principal submatrix $A^{(n-1)}$ has column sums $(0,\dots,0,2\alpha)$. Iterating $n$ times proves $\det A(b) = n!\,\alpha^n$.
\end{proof}

\subsection{Numerical Verification Against \citet{shibata2026dynamic}}
\label{app:verif_table}

Table~\ref{tab:verification} compares all numerical results reported in \citet{shibata2026dynamic} against our exact analytical formulas and \texttt{scipy.linalg.expm}.

\begin{table}[htbp]
\centering
\caption{Comparison of reported numerical results in \citet{shibata2026dynamic} (\textit{J.\ Theor.\ Biol.}\ 637, 112620) with exact analytical formulas and \texttt{scipy.linalg.expm}.}
\label{tab:verification}
\footnotesize
\begin{tabularx}{\textwidth}{@{}>{\raggedright\arraybackslash}p{3.6cm}>{\raggedright\arraybackslash}p{3.6cm}>{\raggedright\arraybackslash}p{4.5cm}>{\raggedright\arraybackslash}X@{}}
\toprule
Quantity & Shibata et al.\ (2026) & Analytical / Exact (\texttt{expm}) & Notes / Mathematical Basis \\
\midrule
Split-dose survival ($0,\ 2.5,\ 23\text{ h}$) & Numerical ODE (Fig.~3)\newline $\mathrm{RSS}=0.300,\ 0.034,\ 0.995$\newline Total $1.329$, $R_{\mathrm{adj}}^2=0.9850$ & Eq.~\eqref{eq:main_split} ($6.53\times 10^{-16}$ vs.\ \texttt{expm})\newline $\mathrm{RSS}=0.2993,\ 0.0341,\ 0.9937$\newline Total $1.3271$, $R_{\mathrm{adj}}^2=0.9858$ & Machine-precision match without ODE solver ($N=9+6+7=22$ pts) \\
Mean inactivation dose $M(0)$ & $3.189\text{ Gy}$ ($\alpha=0.716$)\newline $3.262\text{ Gy}$ ($\alpha=0.7$) & Thm.~B.2: $H_5/\alpha = \frac{137}{60\alpha} = 3.1890\text{ Gy}$ ($3.2619\text{ Gy}$ for $\alpha=0.7$) & Numerical integral reduces to 1-line harmonic number; $3.1934\text{ Gy}$ at $\gamma=7.22$ \\
95\% recovery time $\tau_{95\%}$ & $\approx 8.3\text{ h}$ (Sec.~5.4) & Thm.~C.2: $(\ln 20)/\beta = 502.6\text{ min} = 8.38\text{ h}$ ($8.32\text{ h}$ for $\beta=0.006$) & Universal identity independent of $n, \alpha, D_1$ \\
Single-dose $D_q$ & $2.24\text{ Gy}$ (Table~2) & $\ln(5)/0.716 = 2.2478\text{ Gy}$ & Exact agreement \\
Split-dose $D_q(\tau)$ ($0, 2.5, 23\text{ h}$) & $0\text{ h}$ baseline;\ $1.37\text{ Gy}$ ($2.5\text{ h}$);\ $2.17\text{ Gy}$ ($23\text{ h}$) & Asymptotic (Thm.~C.2): $0.0751, 1.7036, 2.2475\text{ Gy}$\newline Finite fit $[4,10]\text{ Gy}$: $0.0702, 1.6033, 2.1078\text{ Gy}$ & Asymptote is $d\to\infty$; finite-window slope differs by transient $e^{-\mu_k d}$ terms \\
Simultaneous repair ($D_1=5.05\text{ Gy}$) & $7.22\text{ Gy/min}$: $+0.95\%$\newline $1\text{--}5\text{ Gy/min}$: $+6.89\%\text{--}1.38\%$ & Exact \texttt{expm}: $+0.602\%$ at $7.22\text{ Gy/min}$\newline $+4.362\%\text{--}0.870\%$ at $1\text{--}5\text{ Gy/min}$ & Both scale as $1/\gamma$ (Thm.~B.1); $\sim 1.58\times$ factor from evaluation protocol/rounding \\
\bottomrule
\end{tabularx}
\end{table}

% =========================================================================
\section{Spectral Analysis, Eigenvalue Interlacing, WAM Correspondence, and Mean Inactivation Dose}
\label{app:spectral}
% =========================================================================

\begin{table}[tbp]
\centering
\caption{One-to-one correspondence between the 2015 WAM model (one-stage competition \citep{manabe2015whack}) and the $n$-target Shibata model analysis.}
\label{tab:wam_link}
\small
\begin{tabular}{@{}lll@{}}
\toprule
Physical Meaning & 2015 WAM Model (JPSJ) & Present Shibata Analysis \\
\midrule
Repair / clearance rate per unit time & $b_0\ [\mathrm{h}^{-1}]$ & $\beta\ [\mathrm{min}^{-1}]$ \\
Damage transition rate per unit dose & $b_1\ [\mathrm{Gy}^{-1}]$ & $\alpha\ [\mathrm{Gy}^{-1}]$ \\
Characteristic dose rate & $d_* = b_0/b_1$ & $\gamma_* = \beta/\alpha$ \\
Single-unit relaxation eigenvalue & $b_0 + b_1 d$ & $\beta + \alpha\gamma$ \\
Steady equilibrium fraction & $F_{\mathrm{ind}}(\infty)=\dfrac{a_1}{b_1}\cdot\dfrac{d}{d+d_*}$ & $q(\gamma)=\dfrac{\gamma}{\gamma+\gamma_*}$ \\
\bottomrule
\multicolumn{3}{@{}l@{}}{\scriptsize Note: Spontaneous mutation rate $a_0$ of WAM (2015) (analogous to spontaneous cell death) is set to $0$ here.}
\end{tabular}
\end{table}

\begin{theorem}[Eigenvalue Interlacing, Global Bounds, and HDR/LDR Asymptotics]
\label{thm:spectrum}
Let $\gamma_* \equiv \beta/\alpha$ and $q(\gamma) \equiv \alpha/(\alpha+b) = \gamma/(\gamma+\gamma_*)$. For all $n \ge 2$ and $\gamma > 0$:
\begin{enumerate}
  \item $(k-1)(\alpha+b) < \mu_k < k(\alpha+b)$ ($k=1,\dots,n$).
  \item $\alpha q^{n-1} < \mu_1(\gamma) < n\alpha q^{n-1}$ and $\frac{1}{n\alpha}(1+\gamma_*/\gamma)^{n-1} < D_0(\gamma) < \frac{1}{\alpha}(1+\gamma_*/\gamma)^{n-1}$.
  \item First-order HDR perturbation ($\gamma \gg \gamma_*$): $\mu_1(\gamma) = \alpha - (n-1)\frac{\beta}{\gamma} + O(\beta^2/(\alpha\gamma^2))$ (matching the lower bound to first order).
  \item LDR asymptotics ($\gamma \ll \gamma_*$): $\mu_1(\gamma) \sim n\alpha^n(\gamma/\beta)^{n-1}$ (saturating the upper bound with $c_1(\gamma)\to 1$).
\end{enumerate}
\end{theorem}

\begin{proof}
Consider the $(n+1)\times(n+1)$ non-absorbing generator $\bar{A}$ permitting recovery from state $n$, whose top-left $n\times n$ principal submatrix is $A$. On the product space of $n$ independent two-state targets $A_1=\begin{pmatrix}\alpha & -b \\ -\alpha & b\end{pmatrix}$ (eigenvalues $0, \alpha+b$), the Kronecker sum $A_{\mathrm{full}} = \bigoplus_{r=1}^n A_1$ has eigenvalues $\{k(\alpha+b): 0\le k\le n\}$. The count-lumping projection $L$ ($n+1$ rows, full row rank) satisfies $LA_{\mathrm{full}} = \bar{A}L$, and $\bar{A}$ is likewise symmetrizable via $\bar{W}\bar{A}\bar{W}^{-1}$ to an irreducible real symmetric tridiagonal matrix; hence the spectrum of $\bar{A}$ is exactly $\{0,\alpha+b,\dots,n(\alpha+b)\}$. Because $A_{\mathrm{sym}}$ is a principal submatrix of $\bar{A}_{\mathrm{sym}}$, Cauchy's strict interlacing theorem \citep[Theorem~4.3.28]{horn2013matrix}\citep[Ch.~7]{parlett1998symmetric} gives $(k-1)(\alpha+b) < \mu_k < k(\alpha+b)$. Multiplying the bounds for $k=2,\dots,n$ and substituting into $\det A = n!\,\alpha^n$ proves (2).

(3) At $b=0$, the right and left eigenvectors of $\mu_1^{(0)}=\alpha$ are $\mathbf{v}_R = \mathbf{e}_{n-1}$ and $\mathbf{v}_L = (n,n-1,\dots,2,1)^T$ ($\mathbf{v}_L^T\mathbf{e}_{n-1}=1$). Since $R\mathbf{e}_{n-1} = -(n-1)\mathbf{e}_{n-2} + (n-1)\mathbf{e}_{n-1}$, first-order perturbation gives $\mu_1(b) = \alpha + b\,\mathbf{v}_L^T R\mathbf{v}_R + O(b^2/\alpha) = \alpha - (n-1)b + O(b^2/\alpha)$. (4) As $\alpha/b \to 0$, $A/b$ converges to an upper bidiagonal matrix with diagonal entries $0,1,\dots,n-1$, so $\mu_k \sim (k-1)b$ for $k=2,\dots,n$, and hence $\mu_1 = n!\,\alpha^n/(\mu_2\cdots\mu_n) \sim n\alpha^n/b^{n-1} = n\alpha^n(\gamma/\beta)^{n-1}$.
\end{proof}

\begin{theorem}[Exact Polynomial and Harmonic-Number Formula for Mean Inactivation Dose $M(b)$]
\label{thm:mean_dose}
For any $\alpha > 0$ and $b \ge 0$, $M(b) = \int_0^\infty S(D)\,dD$ is given by
\begin{equation}
M(b) = \mathbf{1}^T A(b)^{-1}\mathbf{p}_0 = \mathrm{tr}\bigl(A(b)^{-1}\bigr) = \sum_{m=0}^{n-1} C_{n,m}\frac{b^m}{\alpha^{m+1}}, \qquad
C_{n,m} = \sum_{i=0}^{n-1-m}\frac{(i+m)!}{i!}\,\frac{(n-i-m-1)!}{(n-i)!},
\label{eq:app_Mpoly}
\end{equation}
so $M(0) = H_n/\alpha$, $M'(0) = (n - H_n)/\alpha^2$, and $M(b) \sim b^{n-1}/(n\alpha^n)$ for large $b$. At $n=5$, $C_{5,m} = (\frac{137}{60}, \frac{163}{60}, \frac{137}{60}, \frac{21}{20}, \frac{1}{5})$.
\end{theorem}

\begin{proof}
Integrating $\mathbf{p}'(D) = -A\mathbf{p}(D)$ over $[0,\infty)$ yields $A\mathbf{x} = \mathbf{e}_0$ ($x_i = \int_0^\infty p^{(i)}(D)\,dD$) and $M(b) = \sum_{i=0}^{n-1} x_i$; by \citet[Ch.~5]{keilson1979markov}, $M(b) = \sum_{k=1}^n \mu_k^{-1} = \mathrm{tr}(A^{-1})$. Summing rows $0$ through $i$ of $A\mathbf{x}=\mathbf{e}_0$ gives $(n-i)\alpha x_i - (i+1)bx_{i+1} = 1$ ($0\le i\le n-2$) and $\alpha x_{n-1} = 1$. Solving recursively downward from $i=n-1$ yields $x_i = \sum_{m=0}^{n-1-i}\frac{(i+m)!}{i!}\frac{(n-i-m-1)!}{(n-i)!}\frac{b^m}{\alpha^{m+1}}$, and summing over $i$ establishes Eq.~\eqref{eq:app_Mpoly}.
\end{proof}

\begin{figure}[tbp]
\centering
\includegraphics[width=0.96\textwidth]{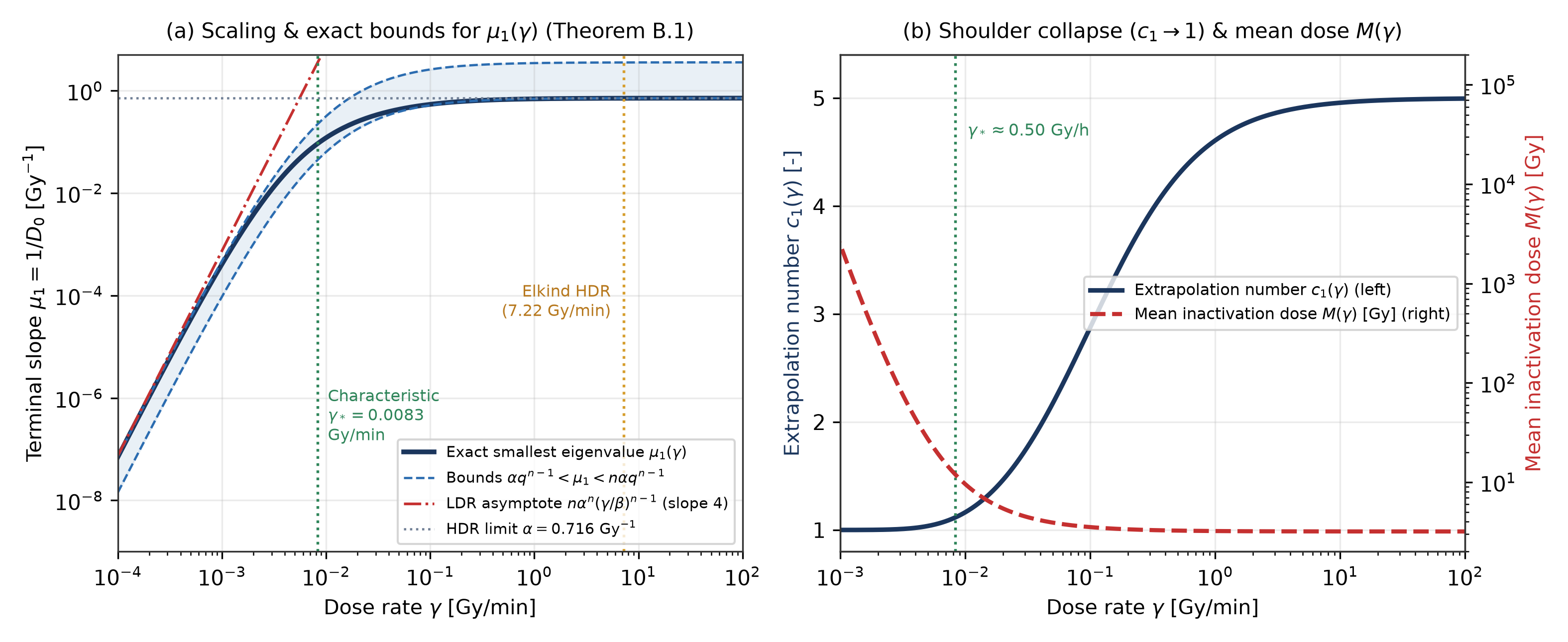}
\caption{(a) All-dose-rate scaling of the smallest eigenvalue $\mu_1(\gamma) = 1/D_0(\gamma)$ for V79 parameters alongside exact Theorem~\ref{thm:spectrum} bounds $(\alpha q^{n-1}, n\alpha q^{n-1})$. (b) Extrapolation number $c_1(\gamma)$ collapsing from $5 \to 1$ and mean inactivation dose $M(\gamma)$ (Theorem~\ref{thm:mean_dose}).}
\label{fig:spectrum}
\end{figure}

% =========================================================================
\section{Closed-Form Split-Dose Solution, \texorpdfstring{$D_q$}{Dq} Identity, and Generalized \texorpdfstring{$G_n(x)$}{Gn(x)} Factor}
\label{app:split_Gn}
% =========================================================================

\begin{lemma}[Recovery Transition Matrix $e^{-\beta\tau R}$]
\label{lem:recovery_matrix}
With $r = e^{-\beta\tau} \in (0,1]$, $(e^{-\beta\tau R})_{k,i} = \binom{i}{k}r^k(1-r)^{i-k}$ for $0\le k\le i\le n-1$ and $0$ for $k > i$.
\end{lemma}

\begin{theorem}[Closed-Form Split-Dose State Distribution and Survival]
\label{thm:split_closed}
Let $q_1 = 1 - e^{-\alpha D_1}$, $q_2 = 1 - e^{-\alpha d}$, and $r = e^{-\beta\tau}$. Post-recovery occupancy ($0 \le k \le n-1$) is $p^{(k)}(T_1+\tau) = \binom{n}{k}(q_1 r)^k(1-q_1 r)^{n-k} - \binom{n}{k}q_1^n r^k(1-r)^{n-k}$, and post-second-dose survival is Eq.~\eqref{eq:main_split}.
\end{theorem}

\begin{proof}
During recovery, each lesion independently remains unrepaired with probability $r$, inducing $\mathrm{Binomial}(i,r)$ transitions among surviving states (Lemma~\ref{lem:recovery_matrix}). Applying Lemma~\ref{lem:recovery_matrix} to $p^{(i)}(T_1) = \binom{n}{i}q_1^i(1-q_1)^{n-i}$, adding and subtracting the $i=n$ term, and applying $\binom{n}{i}\binom{i}{k} = \binom{n}{k}\binom{n-k}{i-k}$ with the binomial theorem yields $p^{(k)}(T_1+\tau)$. Multiplying by second-dose survival probability $1 - q_2^{n-k}$ and summing gives $S = 1 - q_1^n - \sum_{k=0}^{n-1}p^{(k)}(T_1+\tau)q_2^{n-k}$, simplifying to Eq.~\eqref{eq:main_split}.
\end{proof}

\begin{theorem}[Exact Identity Linking $D_q(\tau)$ and Intact Target Pool $U(\tau)$]
\label{thm:Dq_identity}
For any survivor distribution $p^{(i)}(T_1+\tau)$, $S(D_1,\tau,d) = U(T_1+\tau)e^{-\alpha d} + O(e^{-2\alpha d})$ ($\alpha d \gg 1$), yielding Eq.~\eqref{eq:main_Dq} and the universal 95\% recovery time $\tau_{95\%} = (\ln 20)/\beta$.
\end{theorem}

\begin{proof}
Substituting $1 - (1 - e^{-\alpha d})^{n-i} = (n-i)e^{-\alpha d} + O(e^{-2\alpha d})$ into $S(D_1,\tau,d) = \sum_{i=0}^{n-1}p^{(i)}(T_1+\tau)[1-(1-e^{-\alpha d})^{n-i}]$ gives $D_q(\tau) = \alpha^{-1}\ln[U(T_1+\tau)/S(D_1)]$. During recovery, conservation $\sum_{i=0}^{n-1}p^{(i)}(t)=S(D_1)$ and $\frac{dp^{(i)}}{dt}=\beta[(i+1)p^{(i+1)}-ip^{(i)}]$ yield
\[
\frac{dU}{dt} = \sum_{i=0}^{n-1}(n-i)\frac{dp^{(i)}}{dt} = +\beta\sum_{i=0}^{n-1}i\,p^{(i)}(t) = +\beta\bigl[nS(D_1) - U(t)\bigr].
\]
Combining this with $U(T_1) = \sum_{i=0}^{n-1}(n-i)\binom{n}{i}q_1^i(1-q_1)^{n-i} = n e^{-\alpha D_1}$ proves Eq.~\eqref{eq:main_Dq} and $\tau_{95\%} = (\ln 20)/\beta$.
\end{proof}

\begin{figure}[tbp]
\centering
\includegraphics[width=0.96\textwidth]{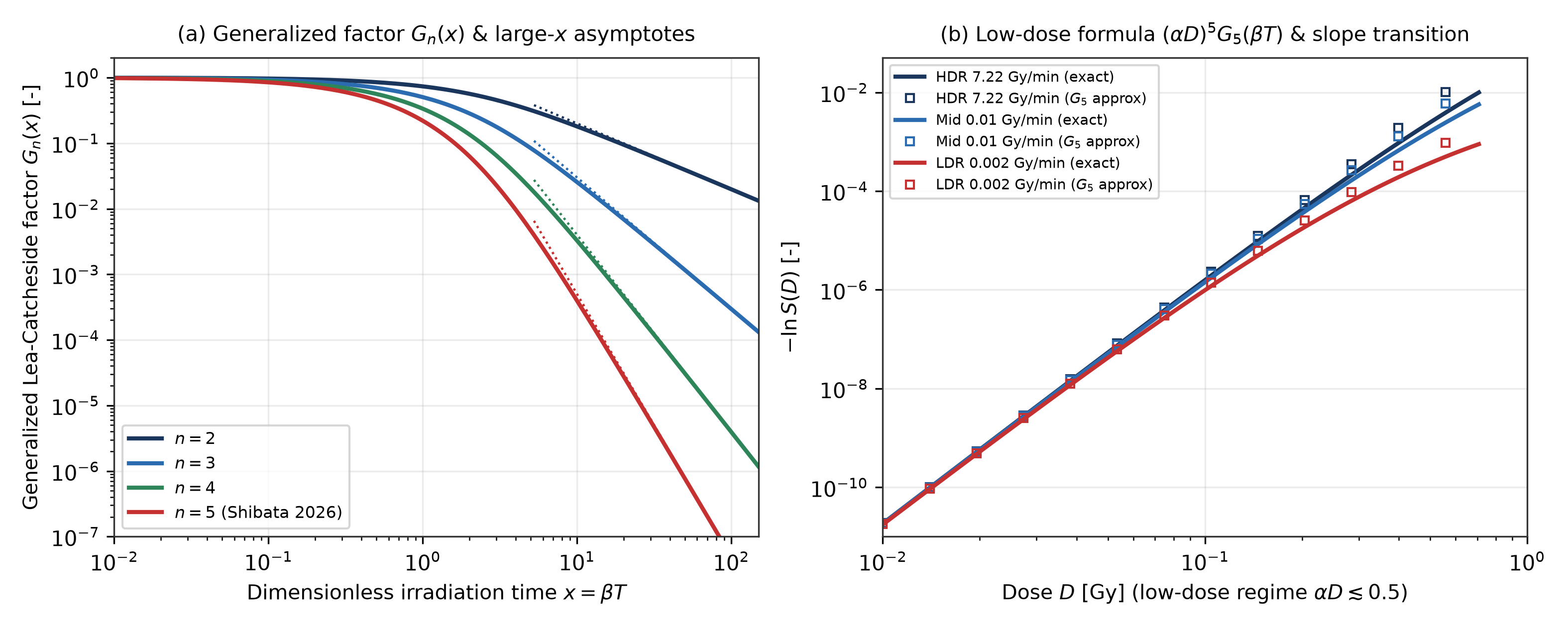}
\caption{(a) Generalized Lea--Catcheside factor $G_n(x)$ for $n=2,3,4,5$ (Theorem~\ref{thm:Gn}) and large-$x$ asymptotes $n/x^{n-1}$. (b) Comparison of exact survival against low-dose approximation $(\alpha D)^5 G_5(\beta T)$ at $n=5$.}
\label{fig:Gn}
\end{figure}

\begin{theorem}[Low-Dose Expansion, Generalized Lea--Catcheside Factor $G_n(x)$, and Exact $n=2$ Solution]
\label{thm:Gn}
For $\mathbf{p}(0)=\mathbf{e}_0$ and constant dose rate $\gamma = D/T$, as $\alpha D \ll 1$:
\begin{equation}
S(D) = 1 - (\alpha D)^n G_n(\beta T) + O\bigl((\alpha D)^{n+1}\bigr), \qquad
G_n(x) \equiv \frac{n}{x^n}\int_0^x (1-e^{-u})^{n-1}\,du.
\label{eq:app_Gn}
\end{equation}
At $n=2$, $G_2(x) = \frac{2}{x^2}(x - 1 + e^{-x}) = G_{\mathrm{Lea\text{--}Catcheside}}(x)$; for $x \ll 1$, $G_n(x) = 1 - \frac{n(n-1)}{2(n+1)}x + O(x^2)$, while for $x \gg 1$, $G_n(x) \sim n/x^{n-1}$ (so $-\ln S \sim n\alpha^n(\gamma/\beta)^{n-1}D$). For $n=2$, the exact all-dose solution for any $b \ge 0$ is
\begin{equation}
S(D) = \frac{\mu_2 e^{-\mu_1 D} - \mu_1 e^{-\mu_2 D}}{\mu_2 - \mu_1}, \qquad
\mu_{1,2} = \frac{3\alpha + b \mp \sqrt{(3\alpha+b)^2 - 8\alpha^2}}{2}, \qquad 1 < c_1 \le 2
\label{eq:app_n2}
\end{equation}
(Fig.~\ref{fig:Gn}).
\end{theorem}

\begin{proof}
Expanding $p^{(i)}(D) = \alpha^i w_i(D) + O(\alpha^{i+1})$ ($w_0=1$) in Eq.~\eqref{eq:main_matrix} yields $\frac{dw_i}{dD} = (n-i+1)w_{i-1} - ib\,w_i$ ($1\le i\le n-1$) and $\frac{dw_n}{dD} = w_{n-1}$. Direct substitution verifies $w_i = \binom{n}{i}y^i$ with $y(D) = (1-e^{-bD})/b$, whence $w_n(D) = n\int_0^D y(D')^{n-1}dD' = D^n G_n(bD)$. At large $x$, $0 \le 1 - (1-e^{-u})^{n-1} \le (n-1)e^{-u}$ yields $G_n(x) = \frac{n}{x^n}(x - O(1)) \sim n/x^{n-1}$. For $n=2$, setting $r = \mu_2/\mu_1$ gives $r + 2 + 1/r = (3\alpha+b)^2/(2\alpha^2) \ge 9/2 \implies r \ge 2$, hence $c_1 = 1 + 1/(r-1) \in (1,2]$.
\end{proof}

% =========================================================================
\section{Monotonicity Theorem for Time-Dependent Recovery \texorpdfstring{$\beta(t)\ge 0$}{beta(t)>=0}}
\label{app:monotonicity}
% =========================================================================

\begin{theorem}[Strict Monotonicity of Survival Under Any Nonnegative Recovery Rate $\beta(t) \ge 0$]
\label{thm:monotonicity}
Under exact ODE evolution $d\mathbf{p}/dt = -\beta(t)R\mathbf{p}$ with any $\beta(t) \ge 0$, the post-recovery state depends solely on the cumulative recovery integral $\Phi_{\mathrm{int}}(\tau) \equiv \int_0^\tau \beta(t)\,dt$ via $\mathbf{p}(T_1+\tau) = \exp(-\Phi_{\mathrm{int}}(\tau)R)\mathbf{p}(T_1)$. Because $\Phi_{\mathrm{int}}(\tau)$ is nondecreasing and post-second-dose survival $S(\tau)$ is \textbf{strictly increasing} in $\Phi_{\mathrm{int}}(\tau)$, nonnegative $\beta(t)$ can never produce a dip.
\end{theorem}

\begin{proof}
Because $R$ is constant, no time ordering is required and $\mathbf{p}(T_1+\tau) = \exp(-\Phi_{\mathrm{int}}(\tau)R)\mathbf{p}(T_1)$. (1) For an acute second dose, setting $r = e^{-\Phi_{\mathrm{int}}} \in (0,1]$ in Eq.~\eqref{eq:main_split} ($0<q_1,q_2<1$) and differentiating gives
\[
\frac{\partial S}{\partial r} = -n q_1(1-q_2)\bigl[q_1 r + q_2(1-q_1 r)\bigr]^{n-1} + n q_1^n(1-q_2)\bigl[r + q_2(1-r)\bigr]^{n-1}.
\]
Rewriting the second term as $q_1[q_1 r + q_1 q_2(1-r)]^{n-1}$ and noting $q_1 q_2(1-r) < q_2(1-q_1 r) \iff q_2(1-q_1) > 0$ yields $\frac{\partial S}{\partial\Phi_{\mathrm{int}}} = -r\frac{\partial S}{\partial r} > 0$. (2) For general dose-rate second exposures, whenever $\Phi_a < \Phi_b$, the binomial thinning in Lemma~\ref{lem:recovery_matrix} induces strict stochastic dominance $\mathbf{p}_b \preceq_{\mathrm{st}} \mathbf{p}_a$ toward lower damage states, guaranteeing $S(\Phi_b) > S(\Phi_a)$ by coupling of birth--death processes.
\end{proof}

% =========================================================================
\section{Proofs of the Four Biophysical Scaling Propositions (Propositions~E.1--E.4)}
\label{app:biophysics}
% =========================================================================

\subsection{Wald's Identity and Target-Size Correction (Section~\ref{subsec:target_arm})}
\label{app:prop_target}

\begin{proposition}[Wald's Identity, Redundant-Hit Factor $H_n$, and Effective Target Size]
\label{prop:target_size}
Starting from $\mathbf{p}(0)=\mathbf{e}_0$ at constant dose rate $\gamma$ ($b\ge 0$), the expected total number of hits sustained before reaching the lethal dose $D_{\mathrm{lethal}}$ is $\mathbb{E}[N_{\mathrm{hits}}^{\mathrm{total}}(b)] = n\alpha M(b)$ (reducing to $n H_n$ at $b=0$), the effective hit efficiency is $\eta_n(0) = 1/H_n$, and the net sensitive pool and per-target length satisfy Eq.~\eqref{eq:main_L1}.
\end{proposition}

\begin{proof}
Let $H(D)$ count all Poisson hits onto the $n$ targets. Adding the hit rate $(n-X(D))\alpha$ onto undamaged targets and the redundant re-hit rate $X(D)\alpha$ onto already-damaged targets gives constant total rate $n\alpha$ independent of state $X(D)$. Thus $H(D) - n\alpha D$ is a martingale. Because $\mathbb{E}[D_{\mathrm{lethal}}] = M(b) < \infty$ ($\mu_1 > 0$) is a valid stopping time, Doob's optional stopping theorem (Wald's identity) yields $\mathbb{E}[H(D_{\mathrm{lethal}})] = n\alpha\,\mathbb{E}[D_{\mathrm{lethal}}] = n\alpha M(b)$. Substituting $M(0) = H_n/\alpha$ (Theorem~\ref{thm:mean_dose}) gives $\mathbb{E}[N_{\mathrm{hits}}^{\mathrm{total}}(0)] = n H_n$. When $n$ targets divide the net sensitive fraction $f_{\mathrm{sens}}$ equally, $\alpha = Y_{\mathrm{DSB}}(f_{\mathrm{sens}}/n) = \rho_{\mathrm{DSB}}L_1$, proving Eq.~\eqref{eq:main_L1}.
\end{proof}

\subsection{Genome-Size Scaling Law (Section~\ref{subsec:genome_scaling})}
\label{app:prop_genome}

\begin{proposition}[Dose-Rate-Dependent Genome-Size Scaling]
\label{prop:genome_scaling}
Across cells sharing $(n\ge 2,\beta)$ with $\alpha(G) = \alpha_0\tilde{G}$ ($\tilde{G}=G/G_0$), we have $\gamma_*(G) = \gamma_*(G_0)/\tilde{G}$ and $\mu_1(G,\gamma) = \alpha_0\tilde{G}\,\Psi_n(\tilde{G}\tilde{\gamma}_0)$ ($\tilde{\gamma}_0=\gamma/\gamma_*(G_0)$), so the terminal-slope ratio $R_\mu = \mu_1(G_A,\gamma)/\mu_1(G_B,\gamma)$ transitions from $(G_A/G_B)^1$ at HDR to $(G_A/G_B)^n$ at LDR.
\end{proposition}

\begin{proof}
Since $\tilde{\gamma}(G) = \gamma/\gamma_*(G) = \tilde{G}\tilde{\gamma}_0$, $\mu_1(G,\gamma) = \alpha(G)\Psi_n(\tilde{\gamma}(G)) = \alpha_0\tilde{G}\,\Psi_n(\tilde{G}\tilde{\gamma}_0)$. At HDR ($\tilde{G}\tilde{\gamma}_0 \gg 1$), Theorem~\ref{thm:spectrum}(3) gives $\Psi_n \to 1$, so $\mu_1 \sim \alpha_0\tilde{G} \propto G^1$; at LDR ($\tilde{G}\tilde{\gamma}_0 \ll 1$), Theorem~\ref{thm:spectrum}(4) gives $\Psi_n(z) \sim n z^{n-1}$, so $\mu_1 \sim n\alpha_0\tilde{G}(\tilde{G}\tilde{\gamma}_0)^{n-1} \propto G^n$, yielding both limits upon taking the ratio.
\end{proof}

\subsection{Apparent Biphasic Recovery, Fisher Harmonic Series, and Linearizing Diagnostic (Section~\ref{sec:finding_biphasic})}
\label{app:prop_biphasic}

\begin{proposition}[Apparent Biphasic Recovery from Logarithmic Concavity and Single-$\beta$ Diagnostic]
\label{prop:apparent_biphasic}
Set $\Lambda(D_1) \equiv S(D_1)/e^{-\alpha D_1} = \sum_{k=0}^{n-1}q_1^k \in (1,n)$. Under a single recovery rate $\beta > 0$: (1) Eq.~\eqref{eq:main_biphasic} holds identically; (2) the apparent instantaneous rate $\beta_{\mathrm{app}}(\tau) \equiv -\frac{d}{d\tau}\ln[1 - f_{D_q}(\tau)]$ decreases strictly from $\beta(\Lambda-1)/\ln\Lambda$ ($\in (1,\frac{n-1}{\ln n})\beta$) at $\tau=0$ to $\beta$ as $\tau\to\infty$, compressing the macroscopic half-time to $\tau_{1/2}^{(D_q)} = \beta^{-1}\ln(1 + \Lambda^{-1/2}) < \beta^{-1}\ln 2$; and (3) the linearizing transform $\ln(n - e^{\alpha D_q(\tau)}) = \ln[n(1-\Lambda^{-1})] - \beta\tau$ holds exactly.
\end{proposition}

\begin{proof}
From $D_q(0)=\alpha^{-1}\ln(n/\Lambda)$ and $D_q(\infty)=\alpha^{-1}\ln n$, $D_q(\infty)-D_q(0)=\alpha^{-1}\ln\Lambda$. Equation~\eqref{eq:main_Dq} gives $\frac{U(T_1+\tau)}{U(T_1)} = \Lambda - (\Lambda-1)e^{-\beta\tau} = 1 + (\Lambda-1)f_U(\tau)$, proving the first identity of Eq.~\eqref{eq:main_biphasic}; strict concavity of $g(u)=\ln[1+(\Lambda-1)u]/\ln\Lambda$ with $g(0)=0,g(1)=1$ ensures $f_{D_q}(\tau) > f_U(\tau)$ for all $\tau > 0$. Setting $a \equiv 1 - \Lambda^{-1} \in (0,1)$, the residual $R(\tau) = 1 - f_{D_q}(\tau) = -\ln(1 - a e^{-\beta\tau})/\ln\Lambda = \sum_{k=1}^\infty \frac{a^k}{k\ln\Lambda}e^{-k\beta\tau}$ has nonnegative weights summing to $-\ln(1-a)/\ln\Lambda = 1$. (2) Defining $t(\tau) \equiv -\ln(1 - ae^{-\beta\tau})$, we have $ae^{-\beta\tau} = 1 - e^{-t}$ and $\beta_{\mathrm{app}}(\tau)/\beta = (e^{t(\tau)}-1)/t(\tau)$, which decreases strictly from $(\Lambda-1)/\ln\Lambda$ to $1$ as $t(\tau)$ falls from $\ln\Lambda$ to $0$. Solving $f_{D_q}(\tau_{1/2}^{(D_q)}) = 1/2$ gives $1 + (\Lambda-1)(1-e^{-\beta\tau}) = \sqrt{\Lambda}$, i.e., $e^{-\beta\tau} = (1+\Lambda^{-1/2})^{-1}$. Finally, rewriting $e^{\alpha D_q(\tau)} = U(T_1+\tau)/S(D_1) = n - n(1-\Lambda^{-1})e^{-\beta\tau}$ yields the linearizing transform.
\end{proof}

\subsection{Two-Track SSB Coincidence and Genome-Length-Invariant Crossover Dose (Section~\ref{sec:finding_multiscale})}
\label{app:prop_multiscale}

\begin{proposition}[Two-Track SSB Coincidence and Genome-Invariant Crossover Dose $D_{\mathrm{cross}}$]
\label{prop:multiscale_ssb}
For combined-strand SSB density $\rho_{\mathrm{SSB}}$, direct DSB density $\rho_{\mathrm{DSB}}$, interaction span $L_c$, and SSB repair rate $\beta_{\mathrm{SSB}}$, the expected two-track DSB yield and crossover dose $D_{\mathrm{cross}}(T)$ at constant dose rate $\gamma = D/T$ satisfy Eqs.~\eqref{eq:main_ssb}--\eqref{eq:main_dcross}.
\end{proposition}

\begin{proof}
Each strand carries SSB density $\rho_{\mathrm{strand}} = \rho_{\mathrm{SSB}}/2$ ($Y_{\mathrm{SSB}} = \rho_{\mathrm{SSB}}L_{\mathrm{tot}}$). For acute irradiation ($T\to 0$), integrating over plus-strand coordinate $x\in[0,L_{\mathrm{tot}}]$ against the opposite-strand coincidence window $[x-L_c, x+L_c]$ (width $2L_c$, sliding window without double counting) yields the expected number of unordered opposite-strand pairs:
\[
\langle N_{\mathrm{DSB}}^{\mathrm{2\text{-}track}}(D,0)\rangle = \int_0^{L_{\mathrm{tot}}}(\rho_{\mathrm{strand}}D)\,(2L_c\rho_{\mathrm{strand}}D)\,dx = \frac{Y_{\mathrm{SSB}}^2 L_c}{2 L_{\mathrm{tot}}}\,D^2.
\]
For finite duration $T>0$, the probability weight that an earlier SSB at $t_1\in[0,T]$ survives repair at rate $\beta_{\mathrm{SSB}}$ until a second opposite-strand SSB occurs at $t_2\in[t_1,T]$ equals
\[
\frac{2}{T^2}\int_0^T dt_1\int_{t_1}^T e^{-\beta_{\mathrm{SSB}}(t_2-t_1)}\,dt_2 = G_2(\beta_{\mathrm{SSB}}T)
\]
(Theorem~\ref{thm:Gn}), establishing Eq.~\eqref{eq:main_ssb}. Equating this to the one-track direct DSB yield $\rho_{\mathrm{DSB}}L_{\mathrm{tot}}D$ cancels $L_{\mathrm{tot}}$ identically and yields Eq.~\eqref{eq:main_dcross}.
\end{proof}

{\small
\bibliographystyle{plainnat}
\bibliography{references}
}

\end{document}